\documentclass[acmsmall,screen]{acmart}

\AtBeginDocument{%
  }

\setcopyright{acmlicensed}
\copyrightyear{2025}
\acmYear{2025}
\acmDOI{10.1145/3719350}

\acmJournal{TKDD}
\acmVolume{19}
\acmNumber{3}
\acmArticle{78}
\acmMonth{4}

\usepackage[linesnumbered,ruled,vlined]{algorithm2e}

\SetKwInput{KwInput}{Input}                % Set the Input
\SetKwInput{KwOutput}{Output}              % set the Output

\usepackage{multicol,xparse,environ}

\usepackage{graphicx}
\usepackage{amsmath}
\usepackage{booktabs}
\usepackage{longtable}% for long tables
\usepackage{enumitem,mathtools}
\usepackage{mathrsfs, euscript}
\usepackage{indentfirst}

\usepackage{array}
\usepackage{xcolor}
\usepackage{multirow,multicol}
\usepackage{subcaption}

\NewEnviron{auxmulticols}[1]{%
  \ifnum#1<2\relax% Fewer than 2 columns
    \BODY
  \else% More than 1 column
    \begin{multicols}{#1}
      \BODY
    \end{multicols}%
  \fi
}

\newcommand{\scheme}{Anti-GAN\ }
\newcommand{\schemens}{Anti-GAN}

\begin{document}

% Title

% Your title must be in mixed case, not sentence case.
% That means all verbs (including short verbs like be, is, using,and go),
% nouns, adverbs, adjectives should be capitalized, including both words in hyphenated terms, while
% articles, conjunctions, and prepositions are lower case unless they
% directly follow a colon or long dash
\title{Exploiting Defenses against GAN-Based Feature Inference Attacks in Federated Learning}
\titlenote{Published in ACM Transactions on Knowledge Discovery from Data (TKDD), 2025}

\author{Xinjian Luo}
\affiliation{%
  \institution{National University of Singapore}
  \country{Singapore}
}
\email{xinjian.luo@u.nus.edu}

\author{Xianglong Zhang}
\affiliation{%
  \institution{University of Science and Technology Beijing}
  \city{Beijing}
  \country{China}
}
\email{xlzhang@xs.ustb.edu.cn}

\renewcommand{\shortauthors}{X. Luo and X. Zhang}

\thispagestyle{empty}
\pagestyle{plain}

\begin{abstract}
Federated learning (FL) is a decentralized model training framework that aims to merge isolated data islands while maintaining data privacy. However, recent studies have revealed that Generative Adversarial Network (GAN) based attacks can be employed in FL to learn the distribution of private datasets and reconstruct recognizable images. 
In this paper, we exploit defenses against GAN-based attacks in FL and propose a framework, \schemens, to prevent attackers from learning the real distribution of the victim's data. The core idea of \scheme is to manipulate the visual features of private training images to make them indistinguishable to human eyes even restored by attackers.
Specifically, \scheme projects the private dataset onto a GAN's generator and combines the generated fake images with the actual images to create the training dataset, which is then used for federated model training. The experimental results demonstrate that \scheme is effective in preventing attackers from learning the distribution of private images while causing minimal harm to the accuracy of the federated model.
\end{abstract}






% \input{tex/responseletter-v2}

\newpage

\maketitle

\setcounter{page}{1}

\section{Introduction}\label{sec-intro}
Machine Learning (ML), particularly deep learning, has gained widespread adoption in various real-life applications, such as face recognition and robotics, owing to its impressive performance. 
However, ML algorithms are susceptible to different types of attacks, such as model inversion~\cite{fredrikson2015model, zhao2021exploiting} and membership inference~\cite{shokri2017membership, salem2018ml}, which can jeopardize user privacy and lead to unintended information leakage. 
To overcome these challenges, \textit{federated learning} (FL)~\cite{mcmahan2016communication} has emerged as a promising solution, which enables collaborative model training between a central server and multiple clients while ensuring user-level privacy and reducing data fragmentation. 
Despite the advantages of FL, recent research has identified a new threat to its security: GAN-based feature inference attacks~\cite{hitaj2017deep, wang2019beyond}. 
In contrast to traditional feature inference attacks that exploit record-level privacy leakage to infer individual training images from trained models~\cite{melis2019exploiting, luo2020feature, zhang2020secret, luo2022feature}, GAN-based attacks aim to infer the \textit{group-level} information of private training images, i.e, the data distribution. This presents a critical challenge to existing privacy-preserving {\color{black}mechanisms~\cite{gradient-precode, gradient-transform} in FL, as they are primarily designed to counteract record-level attacks~\cite{zhu2020deep,luo2020feature} and are often insufficient for defending against GAN-based attacks.}
%
%
% This highlights the need for developing novel privacy-preserving techniques that can safeguard group-level privacy in general FL scenarios, especially against GAN-based inference attacks~\cite{hitaj2017deep, wang2019beyond}.
%
Despite its significance, the current research landscape has yet to adequately explore this issue.
%
% It is worth noting that the existing privacy-preserving mechanisms
%
In this paper, we aim to address the vulnerability of FL to GAN-based attacks by developing a robust defense that is capable of mitigating the privacy risks posed by such attacks while effectively preserving the model utility.

% GAN~\cite{goodfellow2014generative} was initially proposed to learn the distribution of training datasets.
% Given its impressive performance in data reconstruction, GAN is naturally adopted in inference attacks on FL.
% %
% For example, \cite{hitaj2017deep} proposed a  GAN-based attack on FL, where an adversarial client locally trains a GAN model to first learn the distribution of the victim's private dataset (e.g., face pictures) based on the shared model gradients and then reconstruct the private images accordingly.  
% %
% \cite{wang2019beyond} further exploited user-level privacy leakage by training a multi-task GAN model in a malicious central server. 
% %
% Although the privacy risks in FL have been extensively discussed in the GAN-related privacy work, there is a lack of studies that explore possible defenses against these attacks.

%
% To our best knowledge, this is also the \textit{first} paper that exploits defenses against GAN-based attacks in FL.

There are three primary challenges to consider when designing defenses against GAN-based attacks in the context of FL.
\emph{First}, traditional cryptographic methods, such as homomorphic encryption~\cite{wu2020privacy, keller2020mp} and secure multiparty computation~\cite{bonawitz2017practical, Mohassel17}, offer limited utility. 
\textcolor{black}{
The reason is that these methods require significant computational resources~\cite{wu2020privacy,zhang2020batchcrypt,truex2019hybrid}, thereby restricting their application scenarios. In addition, encrypting local gradients can inhibit anomaly detection methods designed to counteract poisoning-based attacks~\cite{jiang2024distribution, bagdasaryan2018backdoor}. 
}
\emph{Second}, the effectiveness of noise injection, particularly \textcolor{black}{ techniques like DP~\cite{xiao2010differential, truex2019hybrid,ZhuangSZCLZ024}, is limited in mitigating GAN-based attacks}~\cite{hitaj2017deep,jiang2024distribution}. 
While DP provides theoretical privacy guarantees for individual private images, GANs aim to reconstruct representations of specific classes~\cite{triastcyn2020federated, xiao2018generating}, {\color{black}surpassing the initial scope of DP's privacy guarantee.  
Moreover, recent work~\cite{jiang2024distribution} has demonstrated the ability to extract training data distributions from FL systems protected by DP. 
While extensions of DP to group-level privacy have been explored~\cite{geyer2017differentially,jiang2024protecting}, these approaches often compromise model performance under moderate privacy budgets~\cite{geyer2017differentially}, and the application scenarios are typically limited to simple models like logistic regression or support vector machines~\cite{jiang2024protecting}.}
\emph{Third}, the success of GAN-based attacks is closely tied to the practicality of the federated model and \textcolor{black}{the memorization nature of neural network models}~\cite{carlini2018secret}. In other words, defending against such attacks becomes challenging as long as the victim's local model maintains a high classification accuracy~\cite{hitaj2017deep}.

It is worth noting that existing defense mechanisms, such as cryptographic methods and noise injection, primarily focus on safeguarding privacy during the training phase.
{\color{black}Considering their limited effectiveness in mitigating GAN-based attacks,
we shift our attention for privacy protection from the model training phase to the data generation phase}.
Correspondingly, we propose a defense mechanism called \scheme to protect the group-level information against GAN-based attacks.
The core idea underlying \scheme is to generate training images using a GAN model, where the visual characteristics of the generated images are obfuscated through a carefully designed loss function. This approach ensures that the attacker's GAN model can only learn the distribution of visually indistinguishable images.
However, as the distribution of the obfuscated images may differ from that of real images, this strategy could potentially result in a degradation of accuracy when the federated model is tested on real images.
To overcome this issue, we devise a novel structure for the victim's GAN model, enabling effective preservation of the classification features from real images while reasonably corrupting the visual features of the generated images.
To further enhance the performance of the federated model, we devise a modified Mixup~\cite{zhang2017mixup} method to generate the final training images by mixing the generated fake images with the real images.
\textcolor{black}{
It is important to note that \scheme is specifically designed to obfuscate object details rather than the entire visual features of training images. \scheme can preserve the semantic information necessary for the downstream classifier, such as the outline of a human face, while rendering object details, such as eyes and noses, indistinguishable to the human eye. Consequently, our scheme can significantly reduce the accuracy degradation caused to the federated model and ensure that images reconstructed by attackers are rendered useless.
}

\begin{figure*}[t]
\center
\begin{minipage}[h]{0.99\textwidth}
\centering
\includegraphics[width=\textwidth]{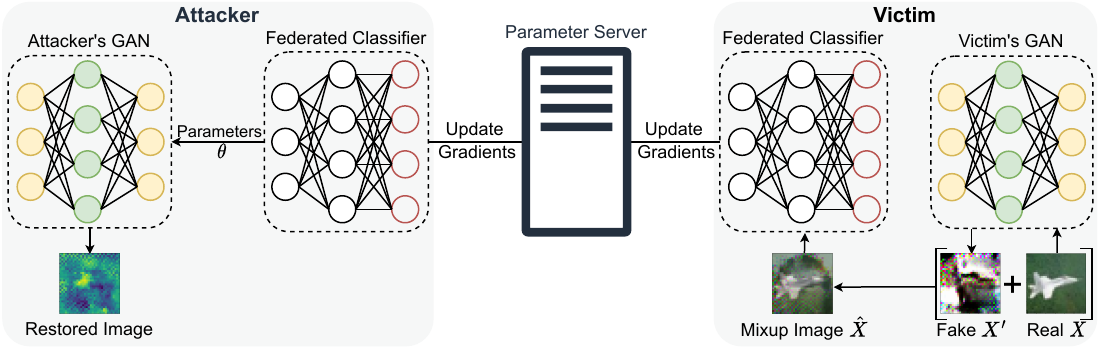}
% \vspace{-2mm}
\caption{The defense framework of \schemens.}
\label{fig-myscheme}
\end{minipage}
% \vspace{-4mm}
\end{figure*}

The framework of \scheme is depicted in Fig.~\ref{fig-myscheme}.
Particularly, the victim initiates the defense process by training a GAN model using the private dataset $X$, generating corresponding fake images $X'$. The objective is to ensure that $X'$ and $X$ possess similar classification features while obscuring the visual features of $X'$. 
Subsequently, the generated $X'$ is mixed with $X$ to form the training dataset $\hat{X}$, which is then used to train the federated model.
To assess the effectiveness of \schemens, we conducted extensive experiments utilizing the MNIST~\cite{mnist}, FashionMNIST~\cite{FashionMNIST}, CelebA~\cite{CelebA}, {and CIFAR10}~\cite{cifar} datasets. The results demonstrate that \scheme successfully reduces the plausibility of images reconstructed by the attacker's GAN model while having minimal impact on the accuracy of the global federated model.
{\color{black}
Our results also show that, beyond countering group-level GAN-based attacks, \scheme can effectively mitigate record-level gradient-matching attacks~\cite{zhu2020deep} as well.
}
Our contributions are summarized as follows:
\begin{itemize}
  \item We explore the defenses against GAN-based feature inference attacks within the context of FL. The proposed defense framework, \schemens, effectively thwarts the attacker's ability to learn distinguishable private visual distribution from the victim's local models.
  \item We introduce an unsupervised learning task to the victim's GAN model, enabling the obfuscation of visual features in the generated images. We design a novel structure for the victim's GAN model to maximally retain the classification features of the real images.
  \item We conduct extensive experiments on four real-world datasets to comprehensively evaluate the performance of \schemens. The results validate the efficacy of our scheme in protecting private images against GAN-based inference attacks.
\end{itemize}

\section{Preliminary}\label{sec-preliminary}
In this section, we introduce the general concepts of federated learning and generative adversarial networks. 

\subsection{Federated Learning}
Federated learning (FL)~\cite{mcmahan2016communication} was initially proposed as a solution to the challenges of data isolation and data privacy protection. It facilitates collaborative training of a shared global model by multiple data holders, referred to as participants or clients, through a multi-round communication process.
The FL workflow entails several key steps. First, all clients agree upon a specific structure for the global model. Subsequently, the server randomly selects a subset of clients and distributes the current state of the global model to each selected client. Finally, each selected client independently trains the model using their local data and exchanges model gradients with other clients through the server.
Suppose $N$ data holders $\{\mathcal{F}_1,\cdots,\mathcal{F}_N\}$ wish to train a shared model $\mathcal{M}_{\text{FED}}$ using their respective datasets $\{\mathcal{D}_1,\cdots,\mathcal{D}_N\}$. The objective of federated learning can be formulated as $\min_{\theta} L(\theta)$ where $L(\theta)=\sum_{k=1}^{N}\frac{n_k}{n}\ell_k(\theta)$.
$n_k=\left|\mathcal{D}_k\right|$ is the size of $\mathcal{F}_k$'s dataset, $n=\sum^{N}_{k=1}n_k$ is the total number of training samples, and $\ell_k(\theta)=\frac{1}{n_k}\sum_{i\in \mathcal{D}_k} \ell(x_i,y_i;\theta)$ is the loss of $\mathcal{F}_k$'s local model. $\ell(x_i,y_i;\theta)$ denotes the loss of prediction on sample point $(x_i,y_i)$ with parameters $\theta$.

\subsection{Generative Adversarial Network}
Generative Adversarial Network (GAN)~\cite{goodfellow2014generative} was initially developed for learning the distribution of training datasets.
A GAN model consists of a generator $G$ and a discriminator $D$. The goal of $G$ is to produce synthetic samples $G(z)$ from random noise $z$ in such a way that $G(z)$ are indistinguishable from real samples $x$ according to the discriminator $D$.
Typically, $D$ and $G$ engage in a minimax  two-player  game such that
% \begin{equation}\label{eq-ganobj}
% \begin{aligned}
  $\min_G\max_D V(D,G) = {} \mathbb{E}_{x\sim p_{data}(x)}[\log D(x)] 
   + \mathbb{E}_{z\sim p_{z}(z)}[\log (1-D(G(z)))]$,
% \end{aligned}
% \end{equation}
where $p_{data}$ denotes the original data distribution, and $p_z$ denotes the distribution of the noise variable $z$. $D(x)$ is the probability output by $D$ that $x$ originates from the real training data rather than being generated by the generator.
Note that the traditional GANs, like DCGAN~\cite{radford2015unsupervised} and WGAN~\cite{gulrajani2017improved}, cannot label the generated images and are thus unsuitable for our context, because the fake images generated by the victim's GAN need to include both the inputs and labels for classification tasks.
Instead, we employ Conditional GAN (CGAN)~\cite{mirza2014conditional} to simultaneously generate images and labels. By feeding real labels $y$ into both $G$ and $D$, CGAN enables control over the classes of the generated images:
\begin{equation}\label{eq-cgan}
\begin{aligned}
  \min_G\max_D V_C(D,G) = {}  \mathbb{E}_{x\sim p_{data}(x)}[\log D(x|y)] 
   + \mathbb{E}_{z\sim p_{z}(z)}[\log (1-D(G(z|y)))].
\end{aligned}
\end{equation}

\section{Problem Statement}\label{sec-problem}

\noindent
\textbf{Threat Model.}
There are two main threat models in GAN-based attacks: one or multiple participants are adversaries (e.g., \cite{hitaj2017deep}), and the central server is the adversary (e.g., \cite{wang2019beyond}). 
Nevertheless, the basic ideas under these attacks are similar, i.e., the attacker leverages the shared model gradients as a discriminator and trains a generator to reconstruct the victim's private data distribution.
Without loss of generality, we focus on the \textit{active} attack model in~\cite{hitaj2017deep}, where a single participant assumes the role of the attacker and can deviate from the training protocol to steal target information. Note that compared to the traditional semi-honest model~\cite{luo2020feature}, the active model provides significantly more benefits to the attacker and therefore presents greater challenges for mitigation.

\vspace{0.8mm}
\noindent
\textbf{Attacking Process.}
% In the context of FL, all participants agree on a shared global model before commencing the training process. 
%
{\color{black} A typical attack scenario used in \cite{hitaj2017deep} is summarized as follows.}
Assume that participant $A$ is the adversary and possesses data with labels $\{b,c\}$, while the victim participant $V$ possesses data with labels $\{a,b\}$. The goal of participant $A$ is to reconstruct the training data distribution belonging to class $a$ owned by $V$.
Note that the label heterogeneity among different participants plays a vital role in motivating GAN-based attacks~\cite{hitaj2017deep,wang2019beyond}. If the adversaries possess samples of the target class in their local datasets, there is no incentive for them to infer the representations of that specific target class.
The attacking process unfolds as follows. First, $V$ trains a local model honestly and uploads the model gradients to the central server. Next, $A$ downloads these shared gradients and utilizes them to update the discriminator of the attack GAN model. Subsequently, $A$ generates a data point with the label $a$ using the GAN and falsely labels it as class $c$.
$A$ proceeds to train the local model using these pseudo-samples and shares the model parameters with $V$, aiming to entice the victim to provide more information about the class $a$. Ultimately,  $A$ can reconstruct an image of class $a$ that is virtually indistinguishable from the samples in participant $V$'s original images.
In this paper, we aim to develop a defense method that can effectively prevent $A$ from reconstructing plausible images from $V$'s shared gradients.

{\color{black}

It is worth noting that, unlike traditional FL scenarios~\cite{mcmahan2016communication}, where participants receive only aggregated global gradients, the two-client scenario used in \cite{hitaj2017deep} allows the attacker to directly receive the victim's gradients. This direct access poses substantial challenges for designing defense methods because, compared to traditional FL scenarios where gradient aggregation can balance out the private information contained in the victim's gradients~\cite{wei2020framework}, the two-client scenario directly exposes the possible private information to attackers. Correspondingly, if our method can effectively mitigate information leakage in the stringent two-client scenario, it can also be adopted in traditional FL scenarios, achieving equal or superior defense effects.

}

\section{\schemens}\label{sec-approach}

In this section, we introduce the proposed defense mechanism, \schemens, for mitigating the data leakage risks posed by GAN-based inference attacks. Note that the victim $V$ is also referred to as the \emph{defender} in this paper, as they are responsible for implementing the defense method.

The main idea of \scheme is to obfuscate the visual features of private training images $X$ that are used in the federated classifier.
One approach is to train a CGAN model using $X$ and generate a corresponding shadow image set $X'$ with visually indistinguishable features. This shadow set $X'$ is then utilized for the federated training process.
However, this straightforward approach encounters two challenges. The first is how to corrupt the visual features of the generated image $X'$. The second is that this method may result in a degradation of accuracy in the federated model, because modifying the visual features of $X$ simultaneously alters its distribution, leading to a potential mismatch between the distributions of the training dataset $X'$ and the original (test) dataset $X$.
To tackle these challenges, we design a three-step approach for \schemens, i.e., obfuscating visual features, preserving classification features, and utilizing Mixup, to generate the final shadow images with desirable properties. 
In following subsections, we provide detailed explanations of each step.

\subsection{Obfuscating Visual Features}
Since the generated samples $X'$ are utilized as the training dataset, the adversary retains the capability to reconstruct representative images of $X'$ using the attack described in~\cite{hitaj2017deep}. Therefore, in order to reduce the distinguishability of the reconstructed images, we need to obfuscate the visual features of $X'$ before initiating the training process.

Note that GAN models are inclined to produce images with sharp details~\cite{isola2017image}. Inspired by PatchShuffle~\cite{shorten2019survey}, we discovered that applying window-wise transformations to an image $x$ can effectively obfuscate its local visual features (e.g., edges and corners) while preserving the overall features (e.g., image patterns). As a result, we design an unsupervised learning task for the defender's generator $G$ to obfuscate the window-wise visual features of the generated images $X'$. Specifically, for each image $x'$ (i.e., $G(z|y)$) generated by $G$, we divide it into $m$ windows of size $s\times s$ and compute the following loss function:
\begin{equation}\label{eq-obf-window}
  \min_G \mathcal{L}_{\text{obf}} \mbox{\quad where \quad} \mathcal{L}_{\text{obf}}=\sum_{i=1}^m \left(Var\left(\boldsymbol{w}_i\left(G(z|y)\right)\right)-v_e\right)^2,
\end{equation}
where $Var(\boldsymbol{w}_i)$ denotes the pixel variance of window $\boldsymbol{w}_i\subseteq G(z|y)$; $v_e$ is the \emph{expected variance}, regulating the level of obfuscation for each window.

\begin{figure}[t!]
\centering
\begin{subfigure}[b]{.45\textwidth}
  \centering
  % include first image
  \includegraphics[width=1\textwidth]{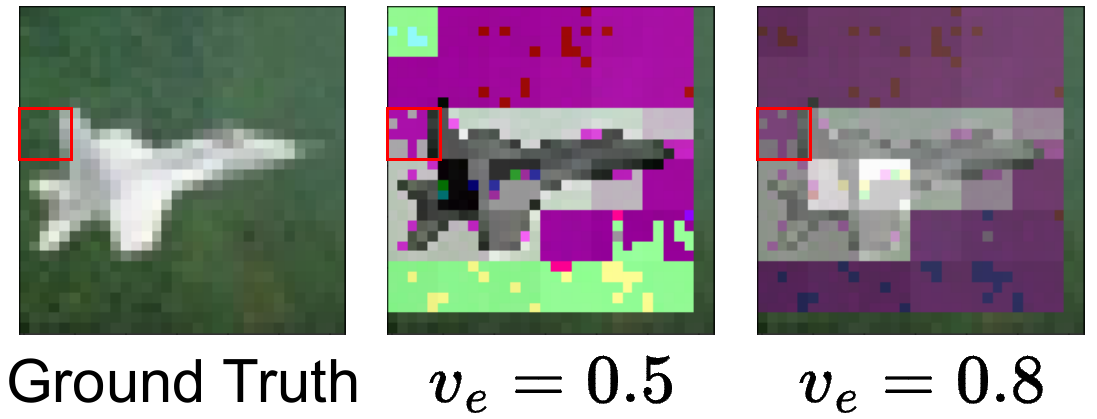}
%   \vspace{-2mm}
  \caption{Images with Different $v_e$}
  \label{fig-var-eg}
\end{subfigure}
\hspace{3mm}
\begin{subfigure}[b]{.412\textwidth}
  \centering
  % include second image
  \includegraphics[width=1\textwidth]{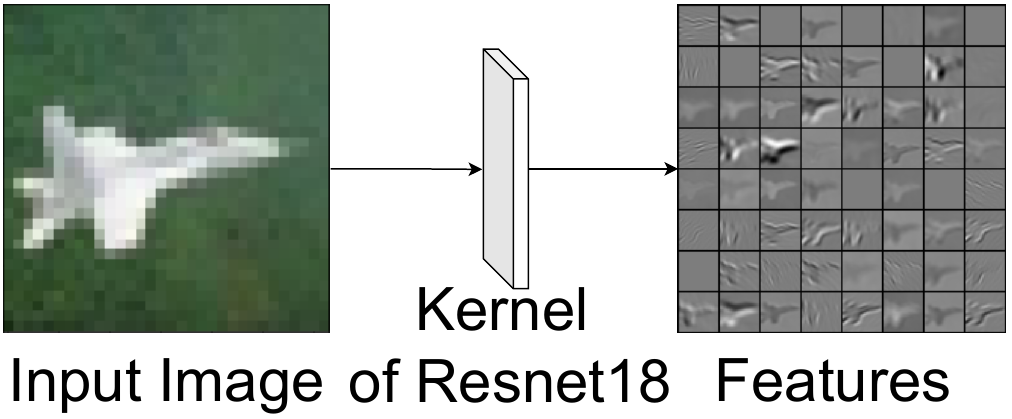}
%   \vspace{-2mm}
  \caption{Classification Features}
  \label{fig-features}
\end{subfigure}
% \vspace{-4mm}
\caption{(a) examples generated from Eq.~\eqref{eq-obf-window}, where the red box denotes a $5\times 5$ window; (b) features extracted by the first convolutional layer of ResNet-18.}
%\vspace{-0.5cm}
\label{fig-two-eggs}
% \vspace{-4mm}
\end{figure}

To demonstrate the effects of $\mathcal{L}_{\text{obf}}$, we employ the gradient descending method to modify the pixel values of an image according to Eq.~\eqref{eq-obf-window} \textit{w.r.t.} different $v_e$. The results are illustrated in Fig.~\ref{fig-var-eg}.
By setting a variance threshold $v_e$ close to the real image variance (e.g., 0.5), the defender can preserve the features of patches containing salient structures, while introducing significant noise to the flat patches, such as the background. Alternatively, by utilizing a larger $v_e$ (e.g., 0.8), increased noise is introduced to obfuscate the entire image, while the overall features are still preserved.

\subsection{Preserving Classification Features}
GAN is used to learn the distribution of the training dataset $X$ via an adversarial process~\cite{goodfellow2014generative}. However, the loss function $\mathcal{L}_{\text{obf}}$ in Eq.~\eqref{eq-obf-window} may hinder the learning process of GAN since it encourages the generator to produce images $X'$ with a different pixel variance $v_e$ than that of $X$. 
This difference can cause a decrease in the accuracy of the federated model tested on $X$ as the distribution of $X'$ may not match that of $X$.
To address this issue, we propose to modify the GAN learning process such that the generated $X'$ have similar classification features to those of $X$.

In the FL training phase, an image $x'$ generated by GAN will be fed into the federated classifier, then this classifier determines the class of $x'$ based on the classification features extracted by a series of convolutional and pooling layers. In other words, to reduce the performance degradation of the federated model after replacing the input $X$ with $X'$, we only need to guarantee that the GAN correctly learns the distribution of the classification features of $X$, instead of the distribution of its visual features. 
Notice that in a deep neural network, the image features extracted by a higher layer (e.g., the final dense layer) are more specific to the task types, whereas the features extracted by a lower layer (e.g., the first convolutional layer) are more general for different tasks~\cite{features}. 
Inspired by this observation, we propose to use the first convolutional layer of ResNet-18~\cite{he2016resnet} pre-trained on ImageNet~\cite{deng2009imagenet} as a \emph{feature extractor} $C$ to extract the low-level image features of $x$, as shown in Fig.~\ref{fig-features}. 
The intuition for the choice of this feature extractor~\footnote{
\textcolor{black}{
In the ablation study of Section~\ref{apdx-exp-ablation}, we replace the first convolutional layer of ResNet-18 with that of DesNet-121~\cite{densenet} as the feature extractor $C$ to evaluate the impact of different choices of $C$ on the defense performance of \schemens. The results demonstrate that the convolutional layers of both ResNet and DenseNet are effective within the framework of \schemens.
}
} is that the ResNet-18 pre-trained on ImageNet is widely used in transfer learning, and its lower convolutional layers can extract extensive features and are general enough for different classification tasks.

\begin{figure}[t!]
% \vspace{-5mm}
\center
\centering
\includegraphics[width=0.85\textwidth]{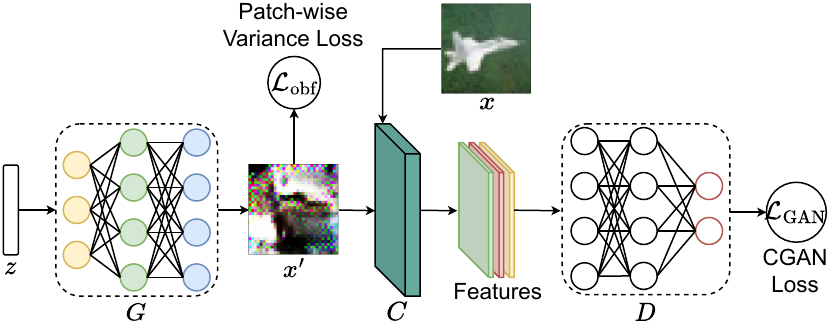}
\caption{The structure of the defender's GAN.}
\label{fig-gan-framework}
% \vspace{-3mm}
\end{figure}

After extracting the classification features $C(x)$ of $x$, we modify the objective function of the defender's GAN as:
\begin{multline}\label{eq-cgan-conv}
  \min_G\max_D \mathcal{L}_{\text{CGAN}} = {} \mathbb{E}_{x\sim p_{data}(x)}[\log D(C(x)|y)]  
   + \mathbb{E}_{z\sim p_{z}(z)}[\log (1-D(C(G(z|y))))],
\end{multline}
where $C(x)$ denotes the features of $x$ extracted by $C$. The discriminator's task is now to distinguish between $C(X)$ and $C(X')$ instead of $X$ and $X'$. In this way, we can ensure that $X'$ has similar classification features to $X$, reducing the performance degradation of the federated model when $X$ is replaced with $X'$.
Finally, the objective function of the defender's GAN is given by
\begin{equation}\label{eq-finalobj}
  G^*(z^*) = arg\min_G\max_D \mathcal{L}_{\text{CGAN}}  
    + \min_G \lambda\mathcal{L}_{\text{obf}},
\end{equation}
where $\lambda$ is a hyperparameter controlling the effect of $\mathcal{L}_{\text{obf}}$.
The structure of the revised GAN model is shown in Fig.~\ref{fig-gan-framework}.

\subsection{Mixup}\label{subsec-mixup}
After training a CGAN model based on Eq.~\eqref{eq-finalobj}, $X'$ can be generated and used as the training images of the federated model by the defender. However, our experiments show that the federated model $f$ trained on $X'$ suffers an accuracy degradation of approximately 20$\%$ when tested on $X$ compared to the model directly trained  on $X$. 
The reason is that the feature extractor $C$ of ResNet-18 has parameters that differ slightly from the parameters of the first convolutional layer $C_f$ of $f$. 
The objective function $\mathcal{L}_{\text{CGAN}}$ can only guarantee that $C(x)\approx C(x')$ instead of $C_f(x)\approx C_f(x')$. This inconsistency between $C_f(x)$ and $C_f(x')$ can lead to the degradation of model accuracy.
To further improve the performance of the federated model trained on $X'$, we use \textit{Mixup} to combine the distributions of $X$ and $X'$.

Mixup~\cite{zhang2017mixup} is a data augmentation method used to enhance the generalization performance of classification models. In Mixup, a training image $\hat{x}$ is generated by mixing two randomly selected private images: $\hat{x}=\mu x_1 + (1-\mu) x_2$, where $\mu$ is a random number sampled from $[0, 1]$. The label $\hat{y}$ is generated accordingly by $\hat{y}=\mu y_1 + (1-\mu) y_2$, where $y_1$ and $y_2$ are one-hot labels.
Considering that the one-hot label $\hat{y}$ can leak the mixup parameter $\mu$~\cite{huang2020instahide,luo2021fusion}, we employ a revised scheme in \scheme to enhance the security of Mixup. For a specific class $c$, we first randomly sample a real image $x$ with label $c$ from $X$ and then generate a fake image $x'$ with label $c$ from the defender's CGAN. After that, the final training image $\hat{x}$ is generated by
% \begin{equation}\label{eq-mixup}
  $\hat{x}=\mu x + (1-\mu) x'$,
% \end{equation}
where $\mu$ is a tunable hyperparameter. The label $\hat{y}$ of $\hat{x}$ is naturally $c$ and leaks no information of $\mu$. 
Correspondingly, the defender's final dataset $\hat{X}$ for FL training is composed of a large number of $\hat{x}$.
Note that to defend against the state-of-the-art attacks on Mixup~\cite{huang2020instahide,luo2021fusion}, we mix each real image $x$ only once, which can effectively mitigate the possible risks~\cite{luo2021fusion}. 
\textcolor{black}{Furthermore, it is worth noting that we combine two images into one training image, as mixing three or more images could lead to model performance degradation and increased computational costs during model training~\cite{zhang2017mixup}.}

\textcolor{black}{To conclude, we summarize the main steps of \scheme in Algorithm~\ref{alg-framework-antigan} for reference.}

\begin{algorithm}[!tb]
\begin{small}
\caption{\textcolor{black}{The overview of \scheme}}
\label{alg-framework-antigan}
\KwInput{the private training dataset $(X,Y)$, the expected pixel variance $v_e$, the feature extractor $C$, the mixup parameter $\mu$, the hyperparameter $\lambda$, a learning rate $\alpha$}
\KwOutput{The generated training dataset $(\hat{X}, \hat{Y})$}

$\boldsymbol{\theta}_{G}$, $\boldsymbol{\theta}_{D}$  $\leftarrow \mathcal{N}(0,1)$ \tcp*{Initialize the generator and discriminator}
\While{\textnormal{$\boldsymbol{\theta}_{G}$ and $\boldsymbol{\theta}_{D}$ have not converged}}  {
    \For{\textnormal{each batch}}{
        $\left\{\left(x^{(i)}, y^{(i)}\right)\right\}_{i=1}^b \gets$ randomly select a batch of samples \;
        $\left\{z^{(i)}\right\}_{i=1}^b \gets$ sample a batch of random vectors \;
        \For{$i \in \{1, \cdots, b\}$}{
            $\left(x^{(i)}\right)' \gets G\left(z^{(i)}, y^{(i)} ; \boldsymbol{\theta}_{G} \right)$ \;
            $\left\{\boldsymbol{w}_j^{(i)}\right\}_{j=1}^m \gets$ divide $\left(x^{(i)}\right)'$ into $m$ windows \;
        }
        $\ell_{\text{obf}}\gets  \frac{1}{b} \sum_{i=1}^b  \mathcal{L}_{\text{obf}}\left( \left\{\boldsymbol{w}_j^{(i)}\right\}_{j=1}^m;v_e \right)$ \tcp*{Obfuscate visual features}
        $\ell_{\text{CGAN}}\gets \frac{1}{b} \sum_{i=1}^b \mathcal{L}_{\text{CGAN}}\left(C\left(x^{(i)}\right), C\left(\left(x^{(i)}\right)'\right), y^{(i)};  \boldsymbol{\theta}_{G}, \boldsymbol{\theta}_{D} \right)$  \;
        $\ell_{\text{G}}\gets \ell_{\text{CGAN}} + \lambda  \ell_{\text{obf}}$
        
        $\boldsymbol{\theta}_{G} \leftarrow \boldsymbol{\theta}_{G} - \alpha \cdot \bigtriangledown_{\boldsymbol{\theta}_{G}}  \ell_{\text{G}} $ \tcp*{Update the generator}
        $\boldsymbol{\theta}_{D} \leftarrow \boldsymbol{\theta}_{D} - \alpha \cdot \bigtriangledown_{\boldsymbol{\theta}_{D}}  \ell_{\text{G}} $  \tcp*{Update the discriminator}
    }
}
$(\hat{X},\hat{Y})\gets \varnothing$ \;
\For{$i \in \{1, \cdots, |X|\}$}{
    $z^{(i)} \gets$ sample a random vector \;
    $\left(x^{(i)}\right)' \gets G\left(z^{(i)}, y^{(i)} ; \boldsymbol{\theta}_{G} \right)$ \;
    $\hat{x}^{(i)}\gets \mu x^{(i)} + (1-\mu) \left(x^{(i)}\right)'$  \tcp*{Mixup}
    $(\hat{X},\hat{Y})\gets (\hat{X},\hat{Y})\cup  \left(  \hat{x}^{(i)},  y^{(i)} \right)  $  \;
}
\Return $(\hat{X}, \hat{Y})$\;
\end{small}
\end{algorithm}

\vspace{0.8mm}\noindent
\noindent
\textbf{Training Methods.} As a game-theoretic framework, the training of Generative Adversarial Networks (GAN) is challenging due to the inherent instability of the optimization process~\cite{goodfellow2014generative}. Without careful parameter tuning, the generator of GAN may fail to converge to a stable point and generate meaningless outputs~\cite{radford2015unsupervised}.
In \schemens, in addition to the difficulty of GAN training, the hyperparameters including $v_e$ in $\mathcal{L}_{\text{obf}}$, the $\lambda$ in the objective function (Eq.~\eqref{eq-finalobj}), and the mixup parameter $\mu$ can all impact the quality of the training dataset $\hat{X}$.
To achieve the desired quality of $\hat{X}$, we propose a two-step training method for \schemens. First, we employ the objective function $\mathcal{L}_{\text{CGAN}}$ and the widely adopted training techniques~\footnote{\url{https://github.com/soumith/ganhacks}} to train a CGAN until convergence. Then, we fine-tune the pre-trained CGAN by optimizing Eq.~\eqref{eq-finalobj} and conduct a grid search to determine the optimal values of the hyperparameters $\{v_e,\lambda,\mu\}$.
Our experimental trials demonstrate that this two-step method is more efficient than directly optimizing Eq.~\eqref{eq-finalobj} from the beginning.

\section{Experiments}\label{sec-exp-res}
In this section, we evaluate \scheme \textit{w.r.t.} the model performance degradation, as well as the defense effectiveness against GAN-based attacks. 
We begin by introducing the experimental setting. Following this, we present the defense performance of \scheme under different parameter settings and compare its performance with DP-based defense methods.

% We first conduct an ablation study to verify the effectiveness of different components of \scheme in Section~\ref{subsec-exp-ablation},  then analyze the impact of different expected variance $v_e$ and Mixup parameter $\mu$ on the model accuracy degradation and attacking results in Section~\ref{subsec-exp-ve} and \ref{subsec-exp-mu}, respectively. 

\subsection{Experimental Setting}\label{subsec-exp-setting}

\noindent\textbf{Datasets.} 
We evaluate the performance of \scheme by conducting experiments on four image datasets: MNIST~\cite{mnist}, FashionMNIST~\cite{FashionMNIST}, CelebA~\cite{CelebA}, \textcolor{black}{and CIFAR10}~\cite{cifar}. 
To ensure fair comparisons, we preprocess all images in these datasets by resizing them to 32$\times$32 pixels and normalizing the pixel values to the range of $[-1, 1]$. In addition, we consider the CelebA dataset for a binary face recognition task, specifically, distinguishing between male and female faces. 

\vspace{0.8mm}
\noindent\textbf{Data Splitting.} 
Regarding data splitting, we initially separate the images into two equal halves based on their classes. One half is allocated to the defender, and the other half is assigned to the attacker, following the approach outlined in \cite{hitaj2017deep}.
It is important to note that this two-client FL setup provides a notable advantage to the attacker due to the direct exposure of the gradients from the other participant.
As a defense mechanism, if \scheme demonstrates effectiveness within this two-client scenario, it inherently implies its viability within more conventional FL setups involving tens or hundreds of participants, because the private information contained in the defender's gradients are naturally balanced out by the gradients contributed by other participants~\cite{wei2020framework}, thus significantly reducing the information available for recovery by the attacker, compared to the two-client setting.

\vspace{0.8mm}
\noindent\textbf{Baseline.} \textcolor{black}{ As discussed in Section~\ref{sec-problem}, GAN-based attacks can be classified into client-side attack~\cite{hitaj2017deep} and server-side attack~\cite{wang2019beyond}. Given the similarity in attack techniques employed by these two types of attacks, the resulting attack outcomes in our experiments are also similar across different settings. Therefore, for clarity, we present only the attack results generated by \cite{hitaj2017deep}.
}
To the best of our knowledge, there is a lack of research dedicated to developing defenses against GAN-based attacks~\cite{hitaj2017deep,wang2019beyond} in the context of FL. 
Most privacy-preserving mechanisms~\cite{gradient-precode, gradient-transform} in FL have been designed to combat gradient-based record-level attacks, making them unsuitable for addressing the challenges posed by GAN-based attacks.
Nevertheless, to showcase the effectiveness of \schemens, we compare its performance with DP-based FL frameworks~\cite{truex2019hybrid, shokri2015privacy}, in which the fundamental privacy-preserving building block is the DP-SGD~\cite{dp-sgd} algorithm, a widely recognized privacy-preserving mechanism with robust theoretical guarantees.
Specifically, in DP-SGD, to achieve $(\epsilon, \delta)$-differential privacy, the defender needs to inject random noises following specific distributions to the model gradients during each training epoch~\cite{dp-sgd}. 
{\color{black}
In addition, we compare \scheme with a Dropout-based defense method~\cite{scheliga2023dropout}, which was designed to counteract gradient-matching attacks~\cite{zhu2020deep} by regularizing the model to prevent it from memorizing private training data.
Through a comparative analysis of accuracy degradation and defensive efficacy among \schemens, the DP-based FL framework~\cite{shokri2015privacy}, and the Dropout-based defense method~\cite{scheliga2023dropout}, we aim to provide insights into the superior privacy-utility trade-off of our proposed approach.
}
%
% In the experiments, we compare the accuracy degradation and defensive performance of \scheme with the DP-based FL framework in \cite{shokri2015privacy} , providing insights into the superior performance of our proposed approach.

\vspace{0.8mm}
\noindent\textbf{Implementation.} 
We now introduce the implementation details of the attack algorithms and FL models in our experiments.
The feature extractor utilized in \scheme is the first convolutional layer of the ResNet-18~\footnote{\url{https://pytorch.org/hub/pytorch_vision_resnet/}} pre-trained on ImageNet.
All neural networks are implemented using the PyTorch~\footnote{\url{https://pytorch.org/}} framework and trained via the Adam optimizer~\cite{adam} with a learning rate of $10^{-4}$. 
DP mechanisms are implemented via the Opacus library ~\cite{opacus}. All experiments are performed on a platform with NVIDIA GTX2060 GPU and AMD R7 4800H CPU.
We employ CGAN as both the defender's and the attacker's GAN models in our study. Detailed structures of the GAN models and the federated classifier are provided in the Appendix for reference.
During the experiments, we observed that variations in the window size specified in Eq.~\eqref{eq-obf-window} and the parameter $\lambda$ in Eq.~\eqref{eq-finalobj} have negligible impact on the quality of the generated images in \schemens. Therefore, for simplicity, we empirically set the window size to $4\times 4$ and $\lambda$ to $250$ for all experiments.
To eliminate potential outliers, we first conduct each experiment ten times independently and then report the averaged results in this paper.

% We evaluate \scheme on three image datasets: MNIST~\cite{mnist}, FashionMNIST~\cite{FashionMNIST}, and CelebA~\cite{CelebA}. 
% %
% To the best of our knowledge, there is a lack of research dedicated to developing defenses against GAN-based attacks in the context of FL. 
% %
% Nevertheless, to showcase the effectiveness of \schemens, we compare its performance with DP-based FL frameworks~\cite{shokri2015privacy}.
% %
% We employ CGAN as both the defender's and the attacker's GAN models in our study. Detailed data splitting, structures of the GAN model and the federated model as well as the FL implementation are provided in the Appendix for reference.
% %
% During the experiments, we observed that variations in the window size specified in Eq.~\eqref{eq-obf-window} and the parameter $\lambda$ in Eq.~\eqref{eq-finalobj} have negligible impact on the quality of the generated images in \schemens. Therefore, for simplicity, we empirically set the window size to $4\times 4$ and $\lambda$ to $250$ for all experiments.
% %
% To eliminate potential outliers, we first conduct each experiment ten times independently and then report the averaged results in this paper.

\vspace{0.8mm}
\noindent\textbf{Metrics.} In our evaluation, we focus on two key aspects of \schemens: the accuracy degradation of the federated model and its defensive performance against GAN-based attacks.
To assess the \textit{accuracy degradation} caused by \schemens, we compute the \textbf{A}ccuracy \textbf{D}egradation \textbf{R}atio (ADR) by $\text{ADR} =  \frac{a_{X} - a_{\hat{X}} }{a_{X}}$,
% \begin{equation}
%     \text{ADR} =  \frac{a_{X} - a_{\hat{X}} }{a_{X}}, 
% \end{equation}
where $a_{\hat{X}}$ denotes the accuracy of the federated model  trained on $\hat{X}$ generated by \scheme and tested on the original images $X$, and $a_{X}$ denotes the accuracy of the model trained and tested on $X$. 
Note that a smaller ADR indicates a lower level of accuracy degradation and better performance of \schemens, and an ADR below $5\%$ is considered acceptable in real-world applications~\cite{huang2020instahide, dp-sgd}.
To evaluate the \textit{defensive performance} of \scheme against GAN-base attacks, we use $\hat{X}$ to train the federated model $f$ and then reconstruct $X$ from $f$ via the attack in~\cite{hitaj2017deep}. The reconstructed images are denoted by $\Tilde{X}$. 
After that, we compare the similarity between $\Tilde{X}$ and $X$ via the mean Structural SIMilarity (SSIM) Index with a window size of $8\times 8$~\cite{wang2004mssim}, which accounts for image luminance, contrast, and structure simultaneously. 
It is important to note that the commonly used $\ell_1$ and $\ell_2$ losses are not applicable in our case. Because these losses serve as record-level similarity metrics, focusing on pixel-level differences between individual images, rather than providing a comprehensive evaluation of group-level similarities.

Since the images in $\Tilde{X}$ are generated based on the distribution of $\hat{X}$ and may not perfectly align with the images in $X$,
we calculate $\text{SSIM}(\Tilde{X}, X)$ as follows.
First, we generate a set $\Tilde{X}$ containing 10,000 labeled images using the attacker's GAN. For each labeled image $(\Tilde{x}_i, \Tilde{y}_i)\in\Tilde{X}$, we compute the average $\text{SSIM}(\Tilde{x}_i, x_j)$ between $\Tilde{x}_i$ and all images $x_j\in X$ with label $y_j=\Tilde{y}_i$. Finally, we calculate $\text{SSIM}(\Tilde{X}, X)$ by averaging the $\text{SSIM}$ values across all $\Tilde{x}_i\in \Tilde{X}$:
\begin{equation}\label{eq-ssim-exp}
    \text{SSIM}(\Tilde{X}, X)=\frac{1}{|\Tilde{X}|}\sum_{\Tilde{x}_i\in \Tilde{X}} \frac{1}{|\{x_j\}|} \sum_{x_j\in X, y_j= \Tilde{y}_i}\text{SSIM}(\Tilde{x}_i, x_j).
\end{equation}
The resulting $\text{SSIM}(\Tilde{X}, X)$ ranges from 0 to 1, where a higher SSIM indicates better visual features in $\Tilde{X}$. Empirically, the generated images $\Tilde{X}$ are indistinguishable to human eyes if $\text{SSIM}(\Tilde{X}, X)<0.3$~\cite{luo2021fusion}.

\subsection{Impact of Different Parameter Settings}\label{subsec-exp-parameter}

\begin{figure*}[t]
\centering
%\captionsetup[subfloat]{captionskip=-0.5mm}
\begin{small}
\begin{tabular}{cccc}
\multicolumn{4}{c}{\hspace{0mm} \includegraphics[height=4mm]{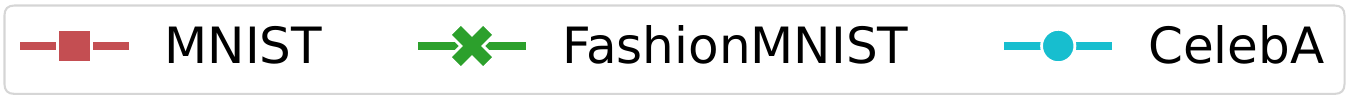}}
\vspace{0mm}  \\
\hspace{-3mm}
\begin{subfigure}[b]{0.25\textwidth}
  \centering
  \includegraphics[width=1\textwidth]{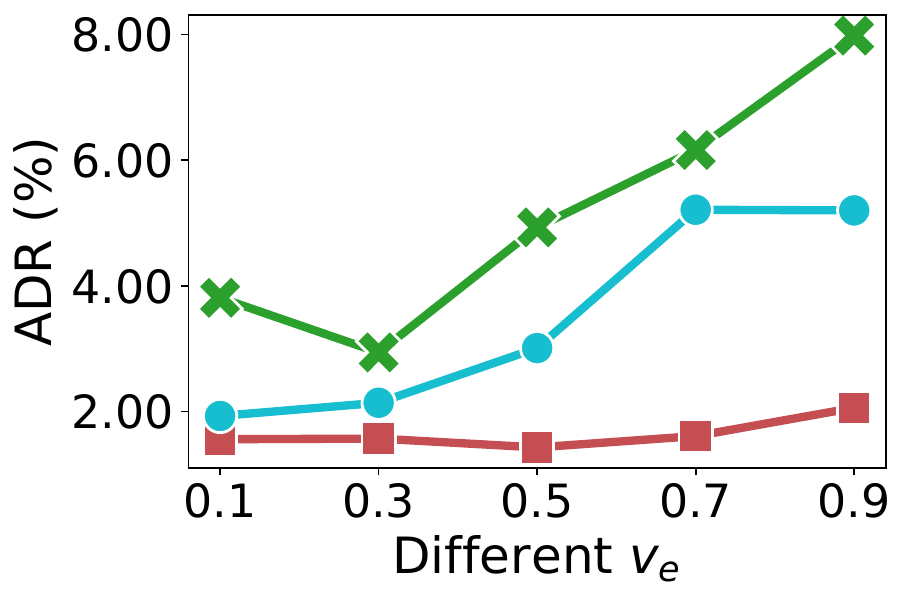}
  \caption{Accuracy Degradation}
  \label{subfig-diff-ve-adr}
\end{subfigure}
&
\hspace{-5mm}
\begin{subfigure}[b]{0.25\textwidth}
  \centering
  \includegraphics[width=1\textwidth]{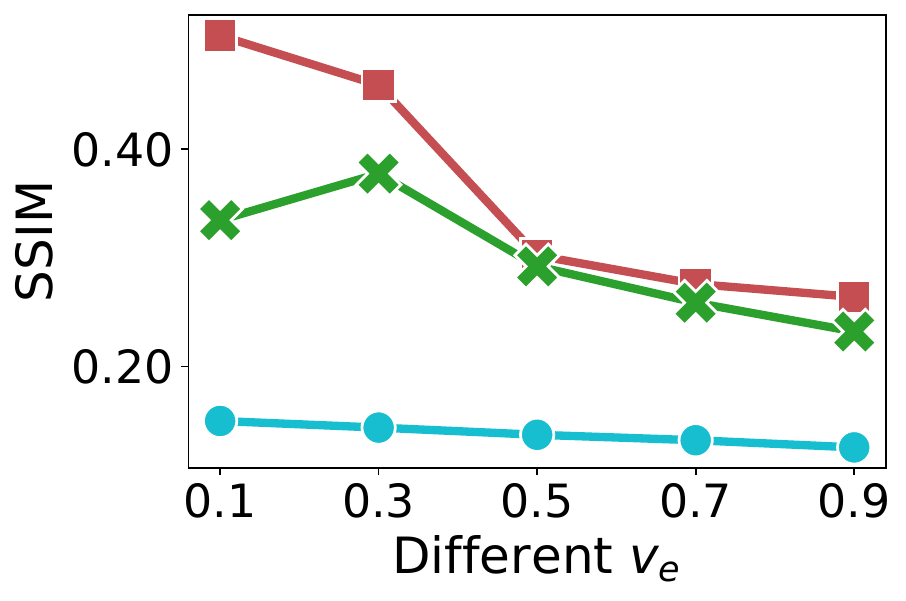}
  \caption{Attack Performance}
  \label{subfig-diff-ve-ssim}
\end{subfigure}
&
\hspace{-5mm}
\begin{subfigure}[b]{0.25\textwidth}
  \centering
  \includegraphics[width=1\textwidth]{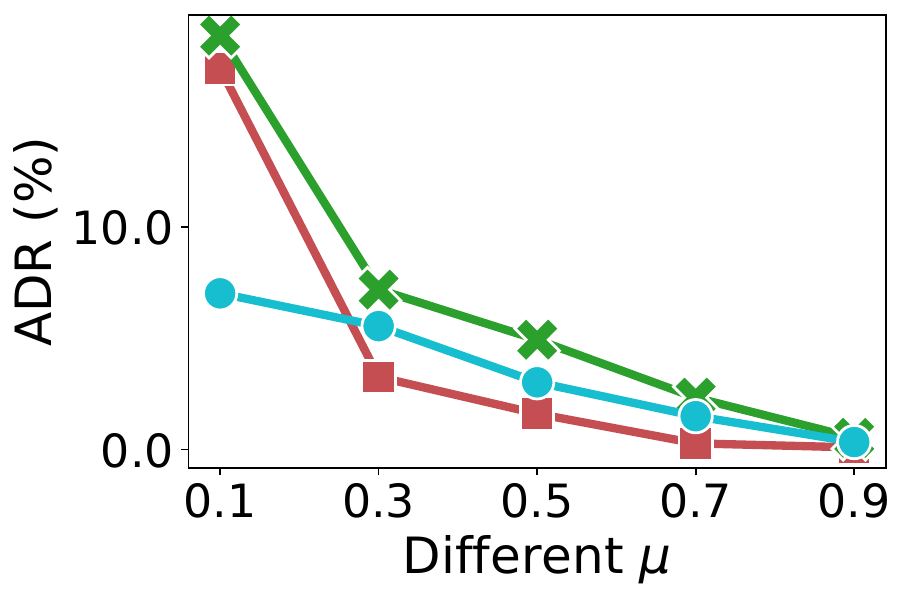}
  \caption{Accuracy Degradation}
  \label{subfig-diff-mu-adr}
\end{subfigure}
&
\hspace{-5mm}
\begin{subfigure}[b]{0.25\textwidth}
  \centering
  \includegraphics[width=1\textwidth]{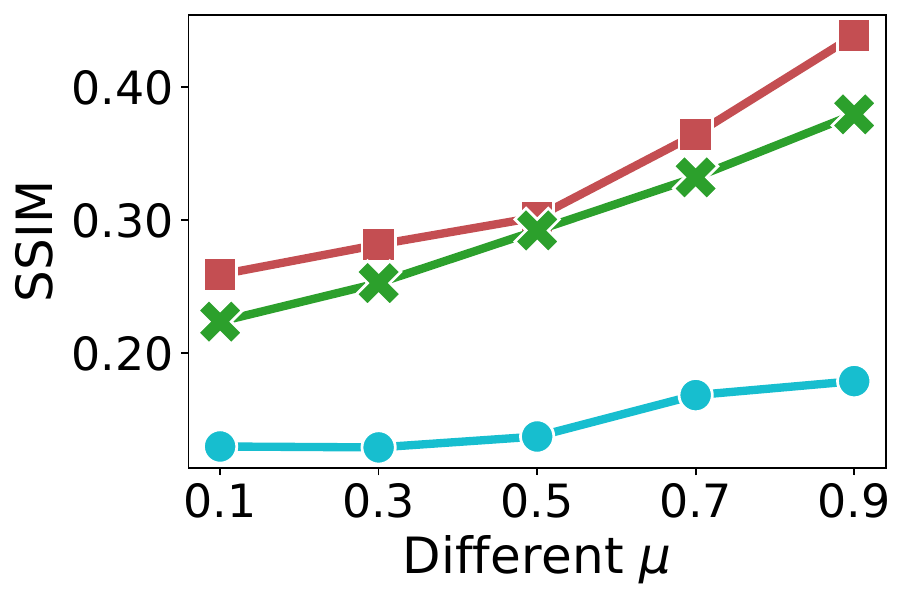}
  \caption{Attack Performance}
  \label{subfig-diff-mu-ssim}
\end{subfigure}
\end{tabular}
% \vspace{-3mm}
\caption{\textcolor{black}{(a)-(b): The accuracy degradation of the federated model and the performance of GAN-based attacks under \scheme with different pixel variances $v_e$ and $\mu=0.5$; (c)-(d): the accuracy degradation and attack performance tested under \scheme with different mixup parameters $\mu$ and $v_e=0.5$.}}
\label{fig-parameter-comparison}
\end{small}
% \vspace{-3mm}
\end{figure*}

\begin{figure*}[t]
% \vspace{-1mm}
\centering
\begin{subfigure}[b]{.47\textwidth}
  \centering
  % include first image
  \includegraphics[width=1\textwidth]{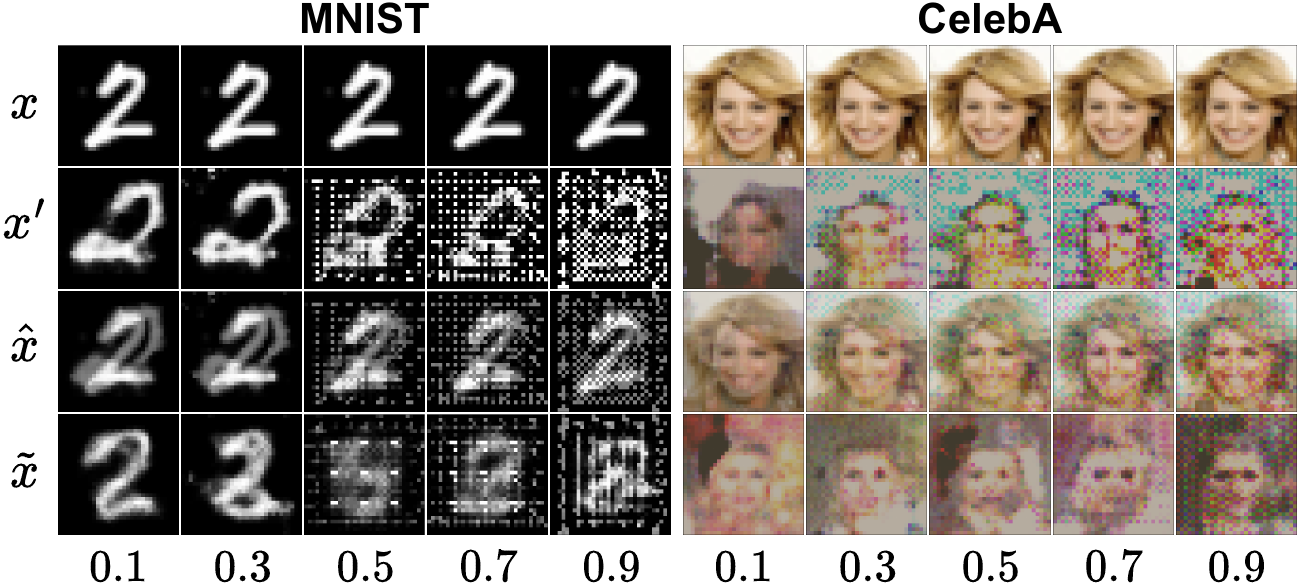}
%   \vspace{-2mm}
  \caption{Different $v_e$ with $\mu=0.5$}
  \label{fig-ve-egg}
\end{subfigure}
\hspace{3mm}
\begin{subfigure}[b]{.47\textwidth}
  \centering
  % include second image
  \includegraphics[width=1\textwidth]{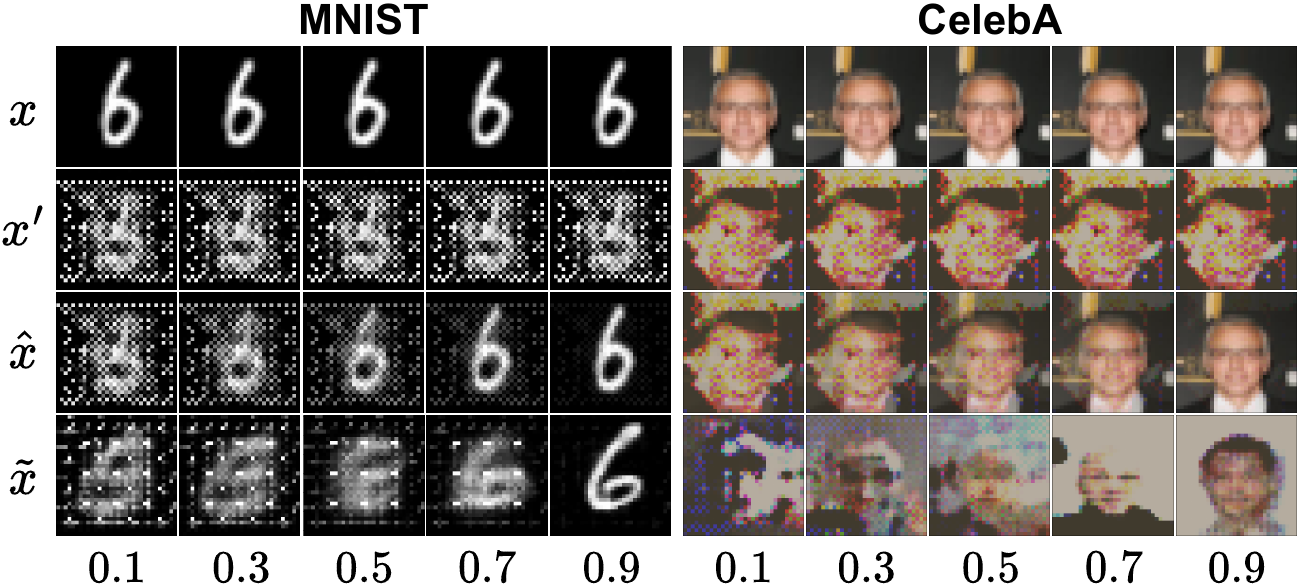}
%   \vspace{-2mm}
  \caption{Different $\mu$ with $v_e=0.5$}
  \label{fig-mu-egg}
\end{subfigure}
% \vspace{-4mm}
\caption{Examples generated with different (a) pixel variance $v_e$ and (b) Mixup parameter $\mu$. $x$ denotes the original images, $x'$ denotes the images generated by the defender's GAN, $\hat{x}$ denotes the Mixup images of $x$ and $x'$, and $\Tilde{x}$ denotes the restored images by \cite{hitaj2017deep}. {\color{black} Note that the images of $\Tilde{x}$ do not correspond to $\hat{x}$ in a one-to-one manner, as GAN models are capable of generating only class representations rather than the original training images.}}
%\vspace{-0.5cm}
\label{fig-two-eggs}
% \vspace{-4mm}
\end{figure*}

% \vspace{0.5mm}
\noindent\textbf{The Expected Pixel Variance $v_e$.}
We firstly investigate the impact of different values of $v_e$ on the performance of \schemens.
Recall that $v_e$ represents the expected pixel variance of the generated images $X'$ and controls the level of visual feature obfuscation in $\mathcal{L}_{\text{obf}}$ (Eq.~\eqref{eq-obf-window}).  
To evaluate the effect of varying $v_e$, we fix the Mixup parameter $\mu=0.5$ and generate different training datasets $\hat{X}$ with $v_e$ values of 0.1, 0.3, 0.5, 0.7, and 0.9. We then assess the corresponding model ADR for the defender and the reconstruction performance for the attacker.
Fig.~\ref{subfig-diff-ve-adr} and \ref{subfig-diff-ve-ssim} show the ADRs and SSIMs tested on \scheme with different $v_e$ values, respectively. 
Additionally, we provide visual examples of images generated by \scheme as well as the attacker's GAN in Figure~\ref{fig-ve-egg}.

The depicted results allow for two key observations.
\textit{First}, the increase in $v_e$ leads to a larger degradation in model accuracy, as depicted in Figure~\ref{subfig-diff-ve-adr}. 
This outcome arises because higher values of $v_e$ drive the distribution of $X'$ further away from the distribution of $X$. Given that the original pixel variance of $X$ is approximately 0.2, such a deviation could significantly alter the distribution of the training images, resulting in reduced testing accuracy of the model.
\textit{Second}, the rise in $v_e$ introduces more noise into the training images $\hat{X}$, thereby diminishing the distinguishability of the images generated by the attacker's GAN, as evident in Figure~\ref{subfig-diff-ve-ssim} and~\ref{fig-ve-egg}.
The presence of additional noise obscures the details of the original images $X$, making them less recoverable by the GAN. Notably, GAN-based reconstruction yields poor visual quality in the restored images $\Tilde{X}$ when $v_e\geq 0.5$, rendering it challenging to discern the edges and corners in MNIST and the facial features in CelebA.
% as demonstrated in Figure~\ref{fig-ve-egg}.
%
Therefore, a privacy-utility trade-off is evident in \scheme as $v_e$ increases. 
Notably, when $v_e\leq 0.5$, the ADRs on the tested datasets remain below $5\%$, while the SSIM values achieved by the attacker generally fall below 0.3 when $v_e\geq 0.5$. This observation suggests that a $v_e$ value of 0.5 strikes the optimal privacy-utility balance.

\vspace{0.8mm}
\noindent\textbf{The Mixup Parameters $\mu$.}
In this section, we assess the impact of the parameter $\mu$ on the performance of \schemens, where $\mu$ controls the amount of information from the real images $X$ incorporated into the training images $\hat{X}$.
To evaluate the effect of $\mu$, we fix $v_e$ at 0.5 and generate different training datasets $\hat{X}$ by varying $\mu$ across the values of 0.1, 0.3, 0.5, 0.7, and 0.9. We then analyze the ADRs of the federated model and the SSIMs achieved by the attacker, as depicted in Figure~\ref{subfig-diff-mu-adr} and \ref{subfig-diff-mu-ssim}, respectively. Additionally, Figure~\ref{fig-mu-egg} showcases visual examples of images generated with different $\mu$ values.
Our observations reveal the existence of a privacy-utility trade-off with respect to $\mu$.
Generally, increasing $\mu$ can reduce the degradation in the accuracy of the federated model, as shown in Figure~\ref{subfig-diff-mu-adr}. This outcome stems from the fact that higher $\mu$ values result in a larger portion of the real images $X$ being incorporated into the training images $\hat{X}$. Consequently, the distribution of $\hat{X}$ becomes closer to the distribution of $X$, leading to improved testing performance.
However, the increase in $\mu$ also provides more advantages to the attacker's GAN model, as demonstrated in Figure~\ref{subfig-diff-mu-ssim}. A larger $\mu$ implies that clearer visual features of $X$ are present in $\hat{X}$, making it easier for the attacker to restore more accurate images and increasing the risk of information leakage.
Moreover, $\mu=0.5$ achieves the best privacy-utility trade-off, as it prevents the attacker from generating distinguishable images ($\text{SSIM}<0.3$) while causing minimal harm to model accuracy ($\text{ADR}<5\%$).

\begin{table*}[t]
\setlength{\tabcolsep}{1pt}
\center
\small
\caption{\textcolor{black}{ Performance comparison between \scheme ($v_e=0.5$, $\mu=0.5$), DP-based FL framework~\cite{shokri2015privacy}, and Dropout-based method~\cite{scheliga2023dropout} }.}\label{tb-comp}
\begin{tabular}{c|c|cccc||cccc}
\hline
\multirow{2}{*}{Methods} & \multirow{2}{*}{Parameters}  & \multicolumn{4}{c||}{ADR} & \multicolumn{4}{c}{SSIM}\\
\cline{3-10}
 &  & MNIST  & F.MNIST & CelebA & \textcolor{black}{CIFAR10} & MNIST  & F.MNIST & CelebA & \textcolor{black}{CIFAR10} \\
\hline 
\hline
\scheme  & /  & ${1.61\boldsymbol{\%}}$ &	${4.93\boldsymbol{\%}}$  & $3.02\%$ & ${8.84{\%}}$ &	{0.3020} & {0.2920}	  &	 {0.1372}  & $0.0933$ \\   
\hline
\multirow{5}{*}{DP-based~\cite{shokri2015privacy}}  & $\epsilon=0.5$    & $6.20\%$ &	$11.94\%$	& $7.69\% $	 & $41.01\%$	  & 0.4515  &	0.2921  & 0.1851 &  0.1122 \\
                            & $\epsilon=1$   & $5.42\%$ 	& $10.22\%$  &	$6.33\%$  & $34.58\%$    &	 0.4893 &	0.3175   &	0.2168  &  0.1219  \\
                         & $\epsilon=2$      & $4.94\%$  &	$8.99\%$ &	$3.46\%$   & $30.38\%$    &	0.5006 &	0.3491	  & 	0.2960 &  0.1472  \\
                         &  $\epsilon=4$     & $3.62\%$  &	$7.96\%$  &	$2.92\%$   & $28.04\%$    &0.5303 &	0.3552   &	0.3245  &  0.1598  \\
                        & $\epsilon=8$       &  $2.32\%$ & $7.39\%$   &  ${2.45{\%}}$   & $20.73\%$ & 0.5365 &	0.3853    &	0.3664  &  0.1540  \\
\hline
 
\multirow{2}{*}{ \textcolor{black}{ Dropout-based~\cite{scheliga2023dropout}} }    & $p=0.1$    & $-0.12\%$ &	$-0.09\%$	& $-0.01\%$	 & $1.24\%$	  & 0.5627  &	0.3962  & 0.3602 & 0.1575  \\
                            & $p=0.5$   & $-0.18\%$ 	& $1.67\%$  &	$1.51\%$  & $3.01\%$    &	 0.5544 &	0.4048   &	0.3796  &   0.1634   \\
\hline
\end{tabular}
% \vspace{-2mm}
\end{table*}

\begin{figure*}[t]
\centering
\begin{subfigure}[b]{.47\textwidth}
  \centering
  % include first image
  \includegraphics[width=1\textwidth]{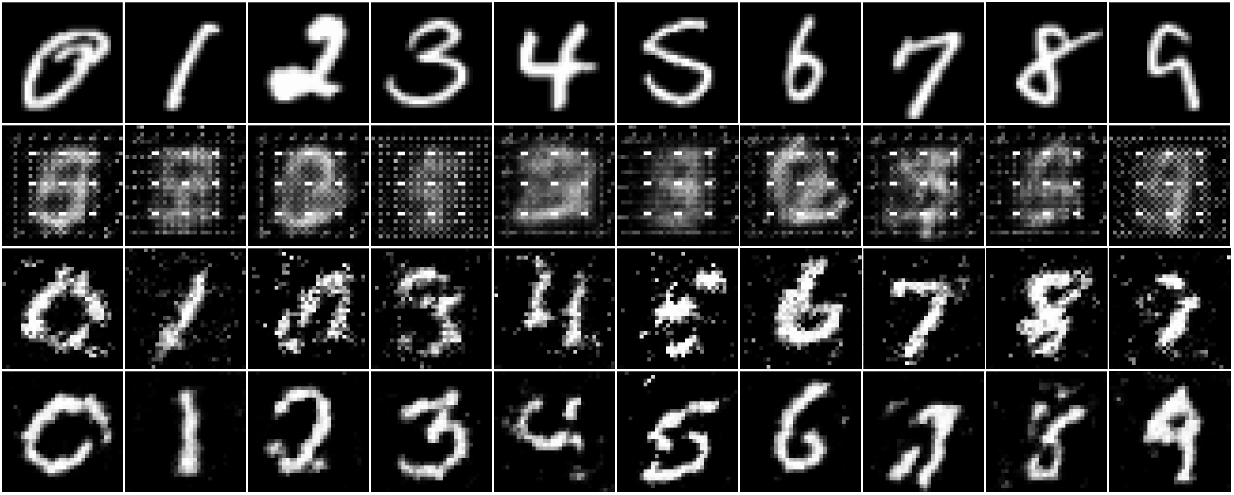}
%   \vspace{-2mm}
  \caption{MNIST}
  \label{fig-dp-mnist}
\end{subfigure}
\hspace{3mm}
\begin{subfigure}[b]{.47\textwidth}
  \centering
  % include second image
  \includegraphics[width=1\textwidth]{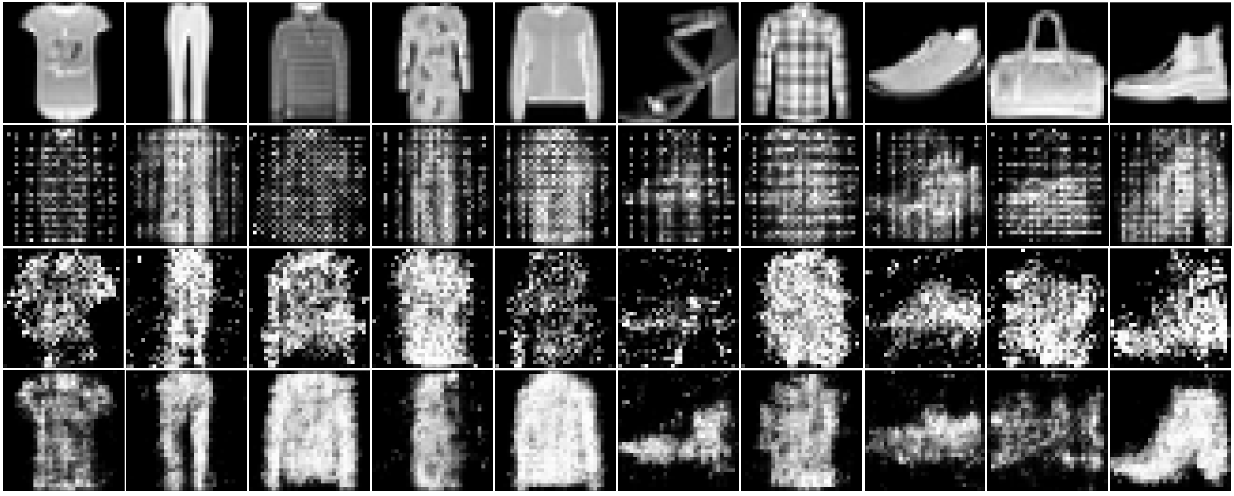}
%   \vspace{-2mm}
  \caption{FashionMNIST}
  \label{fig-dp-fashion}
\end{subfigure}
% \vspace{-2mm}
\caption{The image examples reconstructed by the adversary's GAN model. The first row denotes the original private images, the second row denotes the images reconstructed from \schemens, the third row denotes the images reconstructed from DP FL with $\epsilon=0.5$, and the fourth row denotes the images reconstructed from DP FL with $\epsilon=8$.}
%\vspace{-0.5cm}
\label{fig-comp-dp-two-eggs}
% \vspace{-4mm}
\end{figure*}

\subsection{Comparison with Related Defense Mechanisms}\label{subsec-exp-comp-dp}
In this section, we compare \scheme with the DP-based FL framework~\cite{shokri2015privacy}  \textcolor{black}{and the Dropout-based method~\cite{scheliga2023dropout} } in preserving privacy while maintaining model utility.
For \schemens, we follow the parameter settings suggested in Section~\ref{subsec-exp-parameter}, namely $\mu=0.5$ and $v_e=0.5$. As for DP, we adopt a fixed privacy budget of $\delta=10^{-5}$ as suggested in \cite{dp-sgd} and vary the privacy budget $\epsilon$ across the values of $\{0.5, 1, 2, 4, 8\}$ to train different local models based on $X$. These privacy-preserving models are then utilized as the discriminator in the attacker's GAN to learn the distribution of $X$.
{\color{black}
For the Dropout-based method, we follow the parameter setting in \cite{scheliga2023dropout} and set the dropout rates $p\in \{0.1, 0.5\}$ after each layer of the testing models.
}
Table~\ref{tb-comp} presents the ADRs and SSIMs obtained from \schemens, the DP-based method~\cite{shokri2015privacy}, and the Dropout-based method~\cite{scheliga2023dropout}. Furthermore, Figure~\ref{fig-comp-dp-two-eggs} showcases example images reconstructed using \scheme (the second row), DP FL with $\epsilon=0.5$ (the third row), and DP FL with $\epsilon=8$ (the fourth row).
From these results, we draw three main observations.
First, as shown in Table~\ref{tb-comp}, \textcolor{black}{\scheme incurs the least accuracy degradation in the MNIST, FashionMNIST, and CIFAR10 datasets compared to the DP-based method~\cite{shokri2015privacy}}, while achieving superior defensive performance across all the testing datasets.
Second, while setting $\epsilon=0.5$ in DP FL can provide some mitigation against GAN-based attacks, as depicted in Figure~\ref{fig-dp-fashion}, the defensive performance of DP FL is limited for simple image patterns, such as digital numbers in Figure~\ref{fig-dp-mnist}. Additionally, smaller $\epsilon$ values ($<4$) may lead to accuracy degradation exceeding $5\%$, which could be deemed unacceptable for business companies~\cite{apple-dp}.
{\color{black}
Third, the Dropout-based method~\cite{scheliga2023dropout} incurs minimal accuracy degradation and, under certain conditions, even enhances model performance. This is anticipated as the method in \cite{scheliga2023dropout} can act as a regularization approach that can improve the model's generalization performance. However, its defensive efficacy, as reflected in the SSIM scores for image reconstruction, is the lowest among the three methods. This suggests that the Dropout-based method is less effective in preventing the memorization of training image distributions, thereby offering limited protection against GAN-based attacks~\cite{hitaj2017deep}.
}
In summary, \scheme achieves a more favorable privacy-utility trade-off compared to the DP-based method~\cite{shokri2015privacy} and the Dropout-based method~\cite{scheliga2023dropout}.

\subsection{Defense Performance against Record-Level Attacks}\label{sec-defend-gradient-matching}
Given that \scheme is specifically designed to prevent GAN-based attacks~\cite{hitaj2017deep, wang2019beyond} from retrieving group-level information, an intriguing question arises:  can \scheme effectively counteract attacks at the individual record level? In this section, we evaluate \scheme's defense performance against the gradient-matching attacks~\cite{zhu2020deep}, which aims to reconstruct individual training images via the shared model gradients. 

\begin{figure}[t!]
% \vspace{-5mm}
\center
\centering
\includegraphics[width=0.95\textwidth]{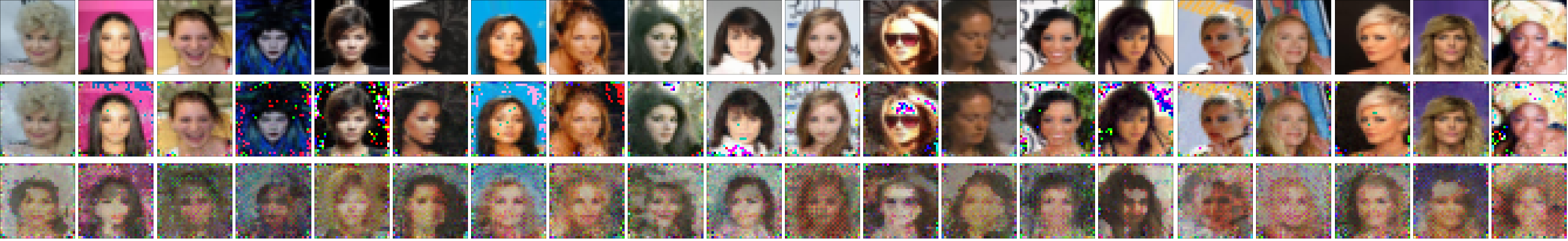}
\caption{The images reconstructed by the gradient matching attack~\cite{zhu2020deep} without (the second row) and with (the third row) \schemens. The first row shows the ground truth images.}
\label{fig-defense-gradient-match}
% \vspace{-3mm}
\end{figure}

Fig.~\ref{fig-defense-gradient-match} shows examples reconstructed by the gradient-matching method proposed in \cite{zhu2020deep}, both without and with the safeguard of \scheme. The images in the second row indicate that gradient-matching attacks can accurately reconstruct original training images at the pixel level. Contrarily, the visual features of the reconstructed images in the third row are obscured, rendering details such as hair and glasses indistinguishable in human faces.
This observation demonstrates \scheme's capability to substantially degrade the performance of record-level gradient-matching attacks in federated learning.
The rationale behind this lies in \scheme's inherent ability to obfuscate visual features in input images. Consequently, while gradient-matching attacks can accurately reconstruct inputs pixel by pixel, the recovered images remain visually blurred.

The findings in Fig.~\ref{fig-defense-gradient-match} underscore that, beyond countering group-level attacks, \scheme can effectively mitigate record-level attacks.

{\color{black}
\subsection{Evaluation of Privacy Leakage on Different Visual Features}\label{subsec-evaluate-visual-features}
In this section, we present experimental results on the CelebA dataset to evaluate whether an attacker can exploit low-quality images generated by \scheme to infer various private visual features. Specifically, we investigate whether \schemens, trained on the gender classification task, can protect the privacy of additional visual features such as hair color and eyeglasses. 
In the experiment, we first train classifiers on real images labeled with various visual attributes including black hair, eyeglasses, smiling, and young. 
These classifiers, once well-trained, are then utilized to predict the labels of images generated by \schemens. 
{\color{black}It is worth noting that classification accuracies are not available in this experiment because \scheme generates images labeled exclusively for the primary gender classification task. As a result, ground-truth labels for other visual features are unavailable, making it impossible to compute classification accuracies for attributes beyond gender.
Instead,} we employ entropy to quantify the uncertainty of model predictions post-softmax~\cite{van2020uncertainty}: $H(p)=-\sum_i p_i(x)\log p_i(x)$, where $p_i(x)$ denotes the probability of $x$ being assigned label $i$ in the model predictions, and higher $H(p)$ values indicate greater uncertainty in the model's predictions. 
The entropy values of model predictions on \scheme for the visual features $\{\text{black hair}, \text{eyeglasses}, \text{smiling}, \text{young}\}$ are $\{0.5846,	0.5574,	0.6086,	0.631\}$, respectively. 
For comparison, the entropy values of model predictions on real images for the four visual features are $\{0.4167,	0.1929,	0.3426,	0.5211
\}$, and the entropy of random guesses is 0.6931.
These results demonstrate that \scheme can substantially increase the uncertainty in model predictions on the generated images with obfuscated visual features, nearly to the level of random guesses, thus significantly mitigating the risks of privacy leakage for various visual features.
}

\subsection{\color{black}Defense Performance under Different Client Settings}\label{sec-exp-client}
{\color{black}
In this section, we evaluate the performance of \scheme under different client configurations within the FL framework. The numbers of clients are chosen from $\{ 2,4,8,{\color{black}16},32\}$. Note that GAN-based attacks~\cite{hitaj2017deep} are effective only in scenarios where the adversary does not possess the target data distribution, which means that the data distribution among different clients in FL should be heterogeneous, non-iid distributions.
In the experiment, we first randomly divide the training data into two halves based on class distribution and similarly partition the clients into two groups. Then, each client in the first group is assigned a specific number of data samples randomly chosen from the first half of the training set, while clients in the second group receive samples from the second half. This approach ensures that the data distributions between the two client groups differ significantly, thereby enabling the attacker in one group to infer the data distribution of a defender in the opposite group using GAN-based attacks. 
To more accurately reflect real-world FL scenarios, we progressively reduce the number of samples per client as the number of clients increases. Specifically, the sample sizes per client for client configurations $\{2, 4, 8, {\color{black}16}, 32\}$ are set to $\{6000, 4000, 2000, 1000, 500\}$, respectively.

Table~\ref{tb-comp-diff-clients} presents the performance of \scheme across different client configurations.
Notably, as the number of clients in the FL framework increases, the degradation in accuracy caused by \scheme diminishes. This trend is due to the reduction in data samples per client as the number of clients rises, which leads to fewer generated samples being used by the defender during training, thereby mitigating the negative impact of \scheme on the performance of the federated model.
Furthermore, the data in Table~\ref{tb-comp-diff-clients} demonstrate that as the number of clients increases, the defense performance of \scheme improves, reflected by a reduction in SSIM scores for image reconstruction. This improvement can be attributed to the limitations of GAN-based attacks~\cite{hitaj2017deep}, which depend on the gradients uploaded by the defender to train the adversarial generator. As the number of samples owned by the defender decreases, the useful information in the uploaded gradients diminishes, thereby hindering the adversary's ability to generate plausible images.
In summary, increasing the number of participating clients in the FL framework can enhance the performance of \schemens.
}

\begin{table*}[t]
\setlength{\tabcolsep}{8pt}
\center
\small
\caption{\textcolor{black}{ Performance comparison under different client settings.} } \label{tb-comp-diff-clients}
\begin{tabular}{c|ccc||ccc}
\hline
\multirow{2}{*}{Client Settings}   & \multicolumn{3}{c||}{ADR} & \multicolumn{3}{c}{SSIM}\\
\cline{2-7}
 & MNIST  & FashionMNIST & CelebA & MNIST  & FashionMNIST & CelebA  \\
\hline 
\hline
 2 Clients     & 1.61\%	&4.93\%	&3.02\%	  & 0.3020&	0.2920&	0.1372  \\
   4 Clients   & 1.19\% &	4.90\%&	3.09\%    &	0.3053&	0.2985&	0.1329    \\
     8 Clients  & 1.08\%	&4.66\%	&3.22\%    &	0.3042	&0.2841	&0.1387   \\
   16 Clients   & 0.98\% &	4.45\%	&2.89\%       &0.2478&	0.2738	&0.1252 \\
    32 Clients  &  0.81\% &	4.54\%	&2.16\% & 0.2345&	0.2506&	0.1203   \\
\hline
\end{tabular}
% \vspace{-2mm}
\end{table*}

\subsection{Ablation Study}\label{apdx-exp-ablation}
\scheme comprises three key components: the loss function $\mathcal{L}_{\text{obf}}$ for obfuscating the visual features of $X'$, the feature extractor for preserving the classification features of $X$ in $X'$, and the Mixup step for enhancing the performance of the federated model. To assess the effectiveness of these components, we remove each component in turn and evaluate the performance of the modified \scheme accordingly.
The MNIST dataset is utilized for testing purposes in this study. It is important to note that the default parameter configuration of \scheme is set to $\mu=0.5$ and $v_e=0.5$.
Table~\ref{tab-ablation} presents the ADR and SSIM metrics, along with sample images of $X$, $X'$, $\hat{X}$, and $\Tilde{X}$ corresponding to different branches of \schemens, from which we make three observations.

First, the model trained without incorporating the $\mathcal{L}_{\text{obf}}$ component within \scheme exhibits the least accuracy degradation. However, the attacker is able to successfully restore the private images with the best visual features. This implies that this modified \scheme provides minimal protection for the defender's private images against GAN-based attacks. This outcome is expected, as the visual features of $X'$ learned by the defender's GAN model are roughly equivalent to those of $X$, thereby offering limited safeguarding for $X$.

Second, in the absence of a feature extractor component, the impact of $\mathcal{L}_{\text{obf}}$ is somewhat neutralized by the learning process of the defender's GAN model. Consequently, the visual features of the generated $X'$ become closer to those of $X$, resulting in a superior attack performance compared to the original \schemens. Concurrently, the defender's GAN primarily focuses on learning the visual features rather than the classification features of $X$, leading to a more substantial degradation in accuracy.

Third, the model trained without the Mixup component experiences the most pronounced accuracy degradation, while the attacker struggles to extract useful information from the training images $\hat{X}$ (i.e., $X'$). As discussed in Section~\ref{subsec-mixup}, the distribution of $X'$ learned by the defender's GAN model can deviate significantly from that of $X$. By incorporating Mixup, which involves the mixing of $X'$ with $X$, we can substantially enhance the model's accuracy, as demonstrated in the original \scheme branch of Table~\ref{tab-ablation}.

{\color{black}
Furthermore, we evaluate the effectiveness of the ResNet feature extractor by substituting it with the first convolutional layer of a pre-trained DesNet-121~\cite{densenet} and then perform experiments on the DesNet feature extractor. The results in Table~\ref{tab-ablation} show that the defense performance of \scheme on the DenseNet feature extractor is comparable to the performance on the ResNet extractor, indicating that the first convolutional layers of various pre-trained deep models can be effectively utilized within the \scheme framework.
}

In summary, the ablation study shows that the original \scheme achieves the optimal privacy-utility trade-off, thereby validating the effectiveness of different components in our design.

\begin{table*}[t!]
\caption{Results of the ablation study.}
\centering
\begin{tabular}{  
>{\centering\arraybackslash} m{1.5cm}  
>{\centering\arraybackslash} m{2.0cm}
>{\centering\arraybackslash} m{2cm}
>{\centering\arraybackslash} m{2cm}
>{\centering\arraybackslash} m{2cm}
>{\centering\arraybackslash} m{2cm}
>{\centering\arraybackslash} m{3cm}
}
\toprule
Branch & The Original \scheme  & No $\mathcal{L}_{\text{obf}}$ & No Feature Extractor $C$ & No Mixup & \textcolor{black}{DenseNet Extractor $C$} \\
\toprule
ADR & 1.61$\%$ & 0.96$\%$ &  1.98$\%$ &  22.55$\%$  & \textcolor{black}{1.45$\%$}\\
% \hline
SSIM & 0.3020 &	0.5680 &	0.3467 &	0.2446 & \textcolor{black}{0.3037}\\
% \midrule
$x$
& \includegraphics[width=0.06\textwidth]{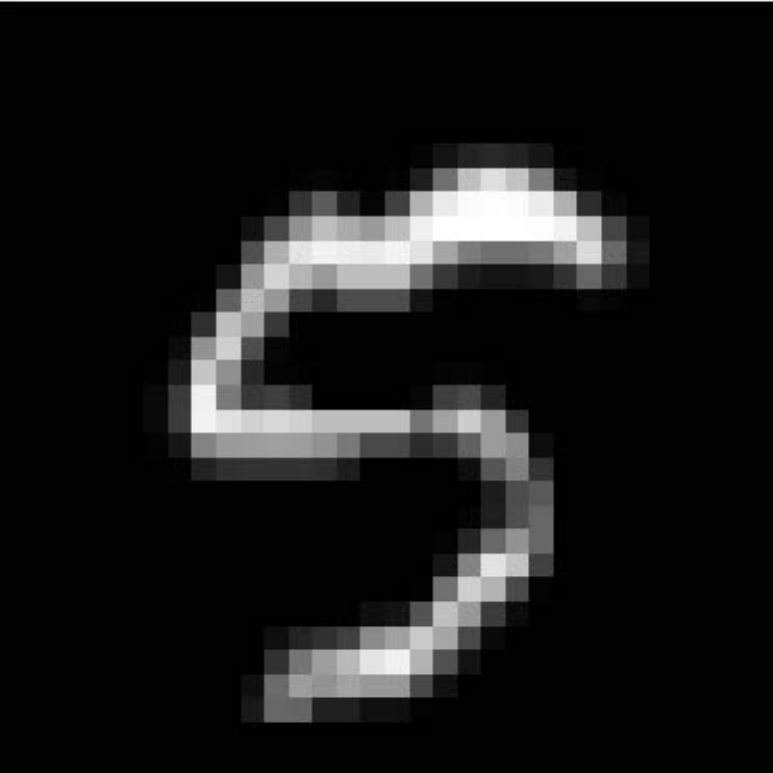} 
& \includegraphics[width=0.06\textwidth]{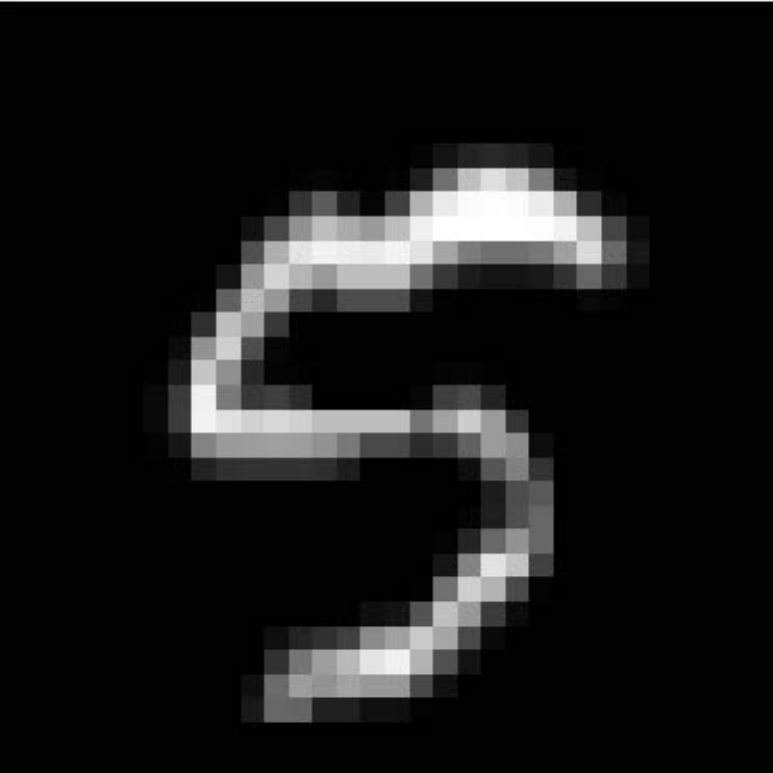} 
& \includegraphics[width=0.06\textwidth]{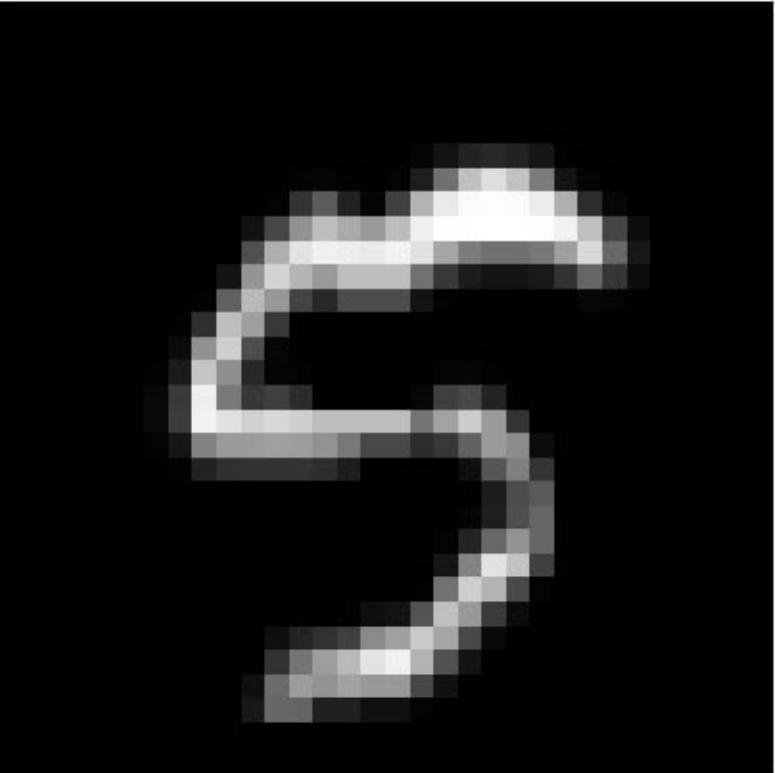} 
& \includegraphics[width=0.06\textwidth]{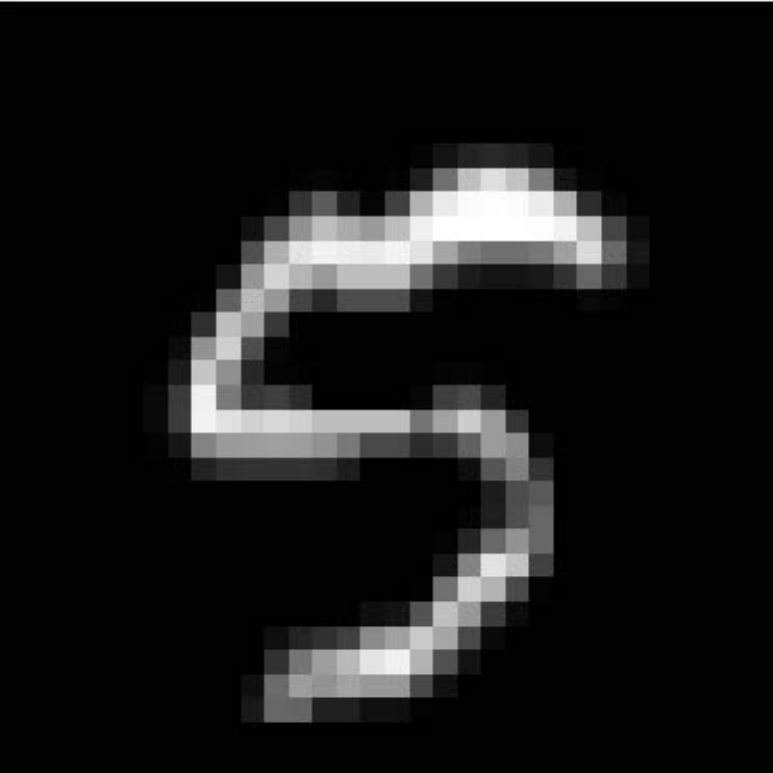} 
& \includegraphics[width=0.06\textwidth]{figures/orig-x.pdf} \\
% \midrule
 $x'$
& \includegraphics[width=0.06\textwidth]{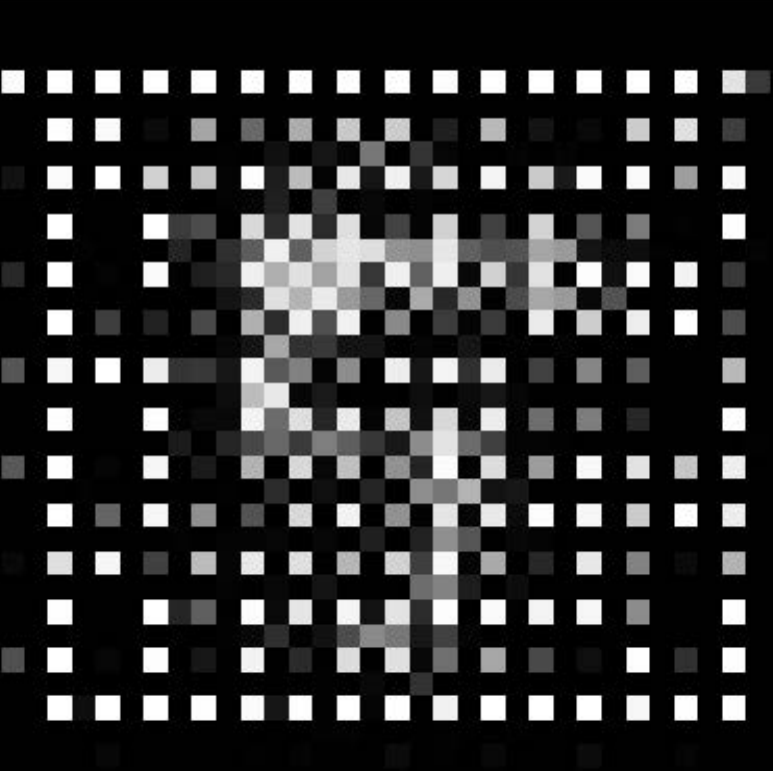} 
& \includegraphics[width=0.06\textwidth]{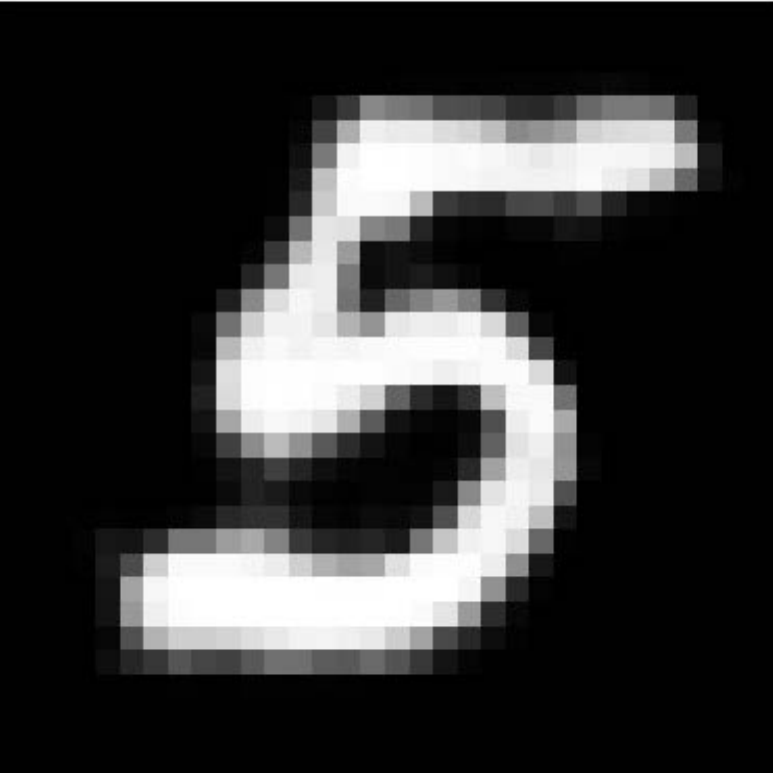} 
& \includegraphics[width=0.06\textwidth]{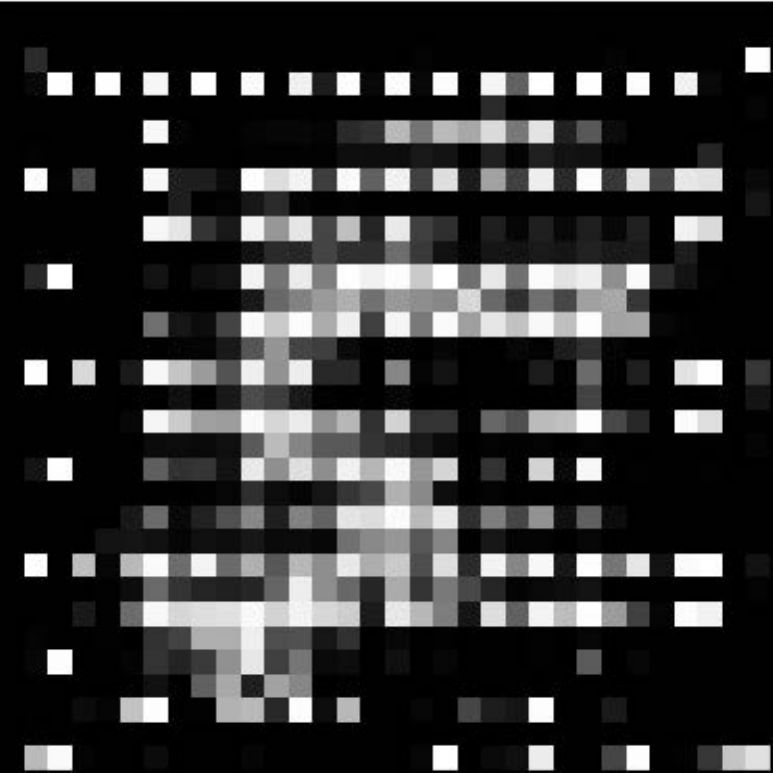} 
& \includegraphics[width=0.06\textwidth]{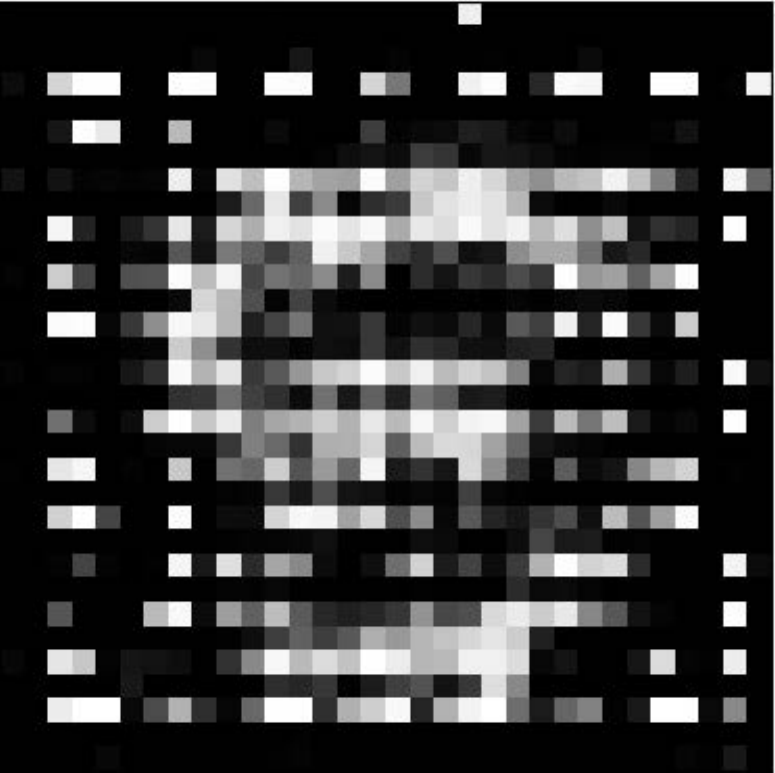}
& \includegraphics[width=0.06\textwidth]{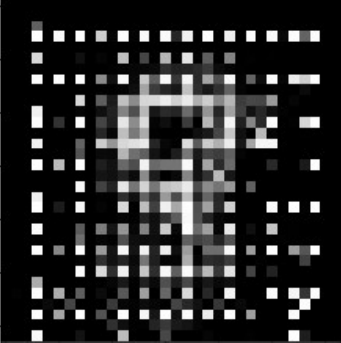}\\
% \midrule
$\hat{x}$
& \includegraphics[width=0.06\textwidth]{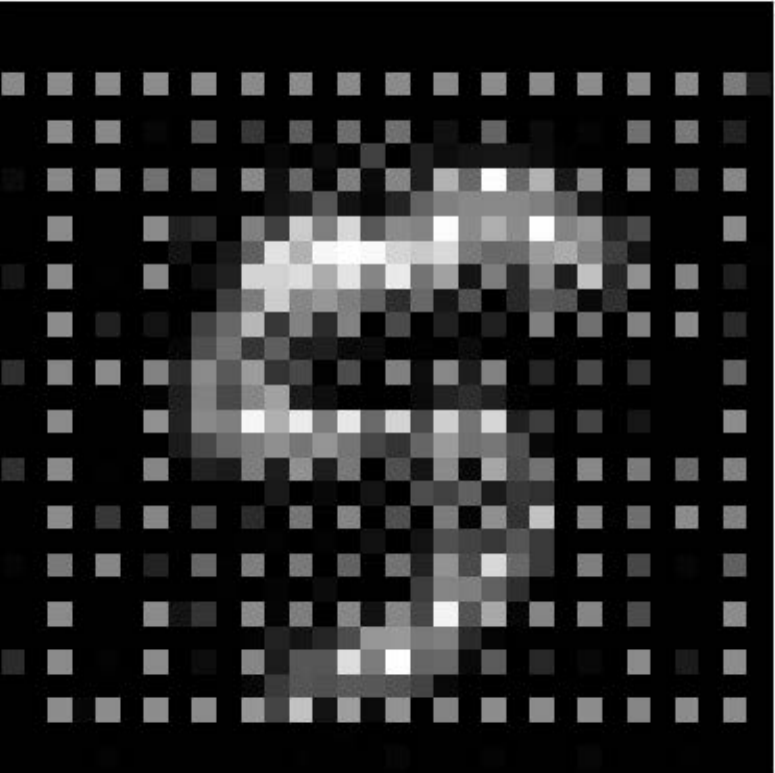} 
& \includegraphics[width=0.06\textwidth]{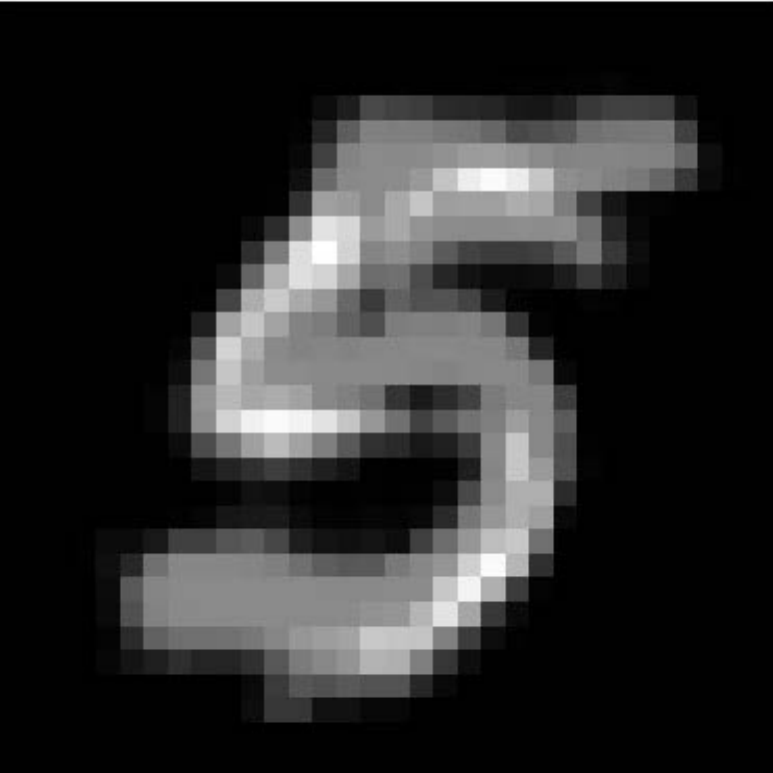} 
& \includegraphics[width=0.06\textwidth]{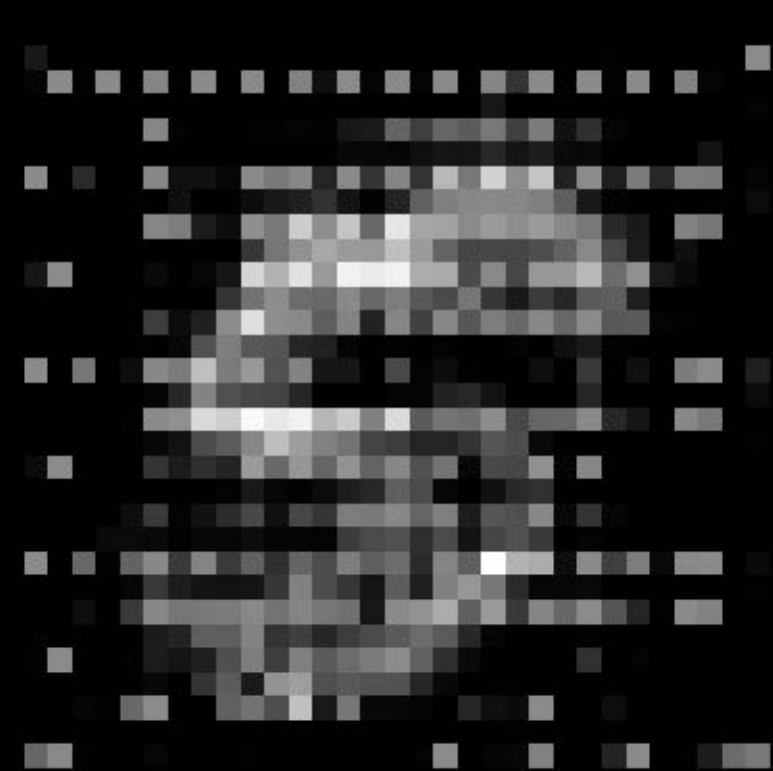} 
& \includegraphics[width=0.06\textwidth]{figures/nomixup-xprime.pdf}
& \includegraphics[width=0.06\textwidth]{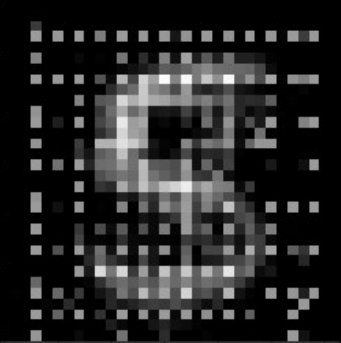} \\
% \midrule
 $\Tilde{x}$
& \includegraphics[width=0.06\textwidth]{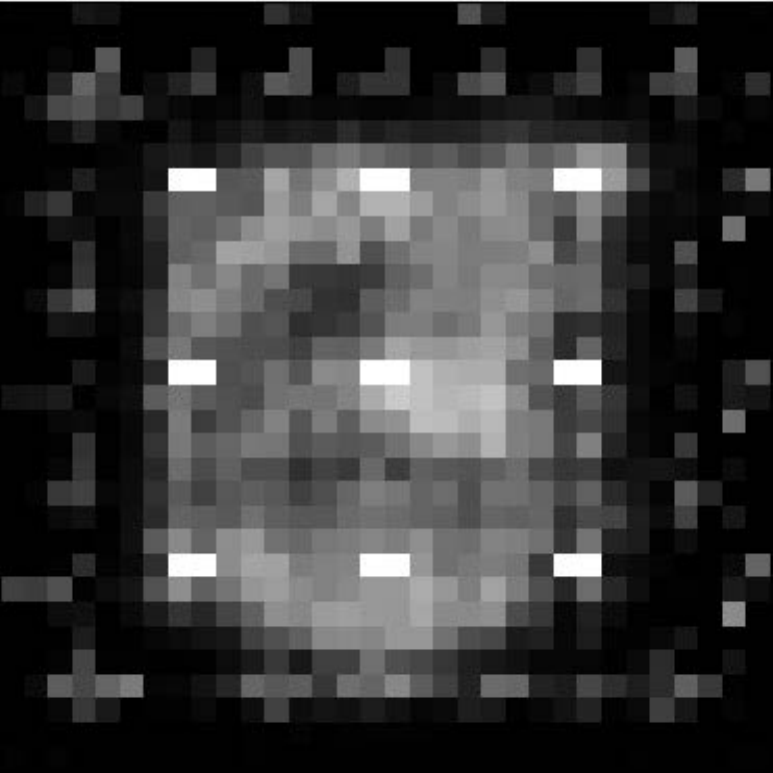} 
& \includegraphics[width=0.06\textwidth]{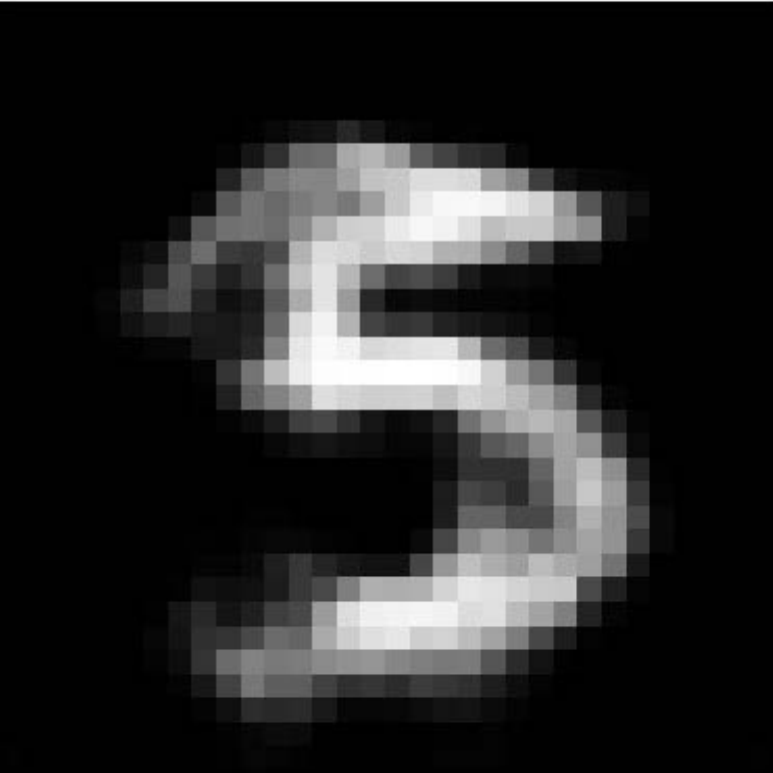} 
& \includegraphics[width=0.06\textwidth]{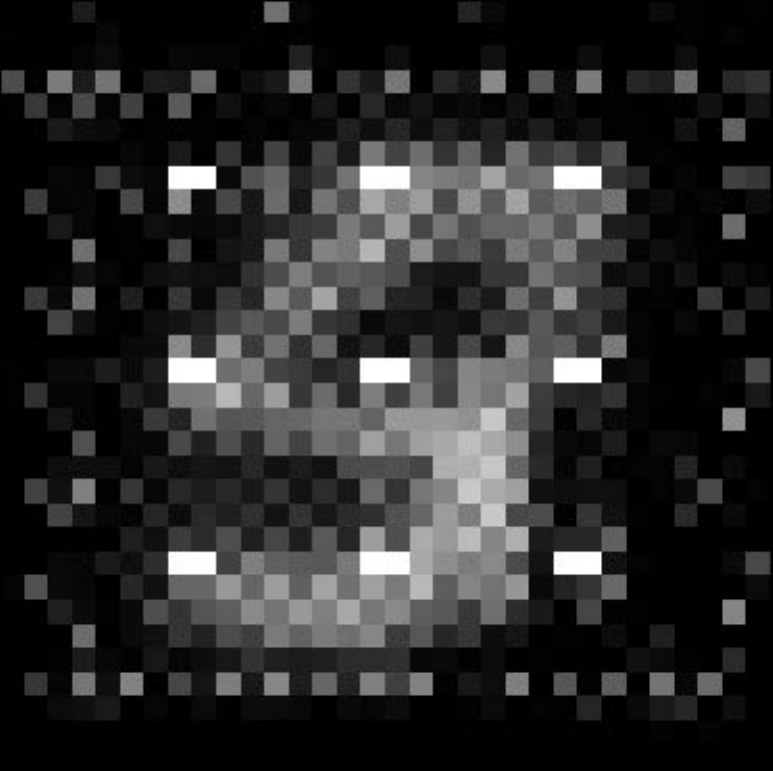} 
& \includegraphics[width=0.06\textwidth]{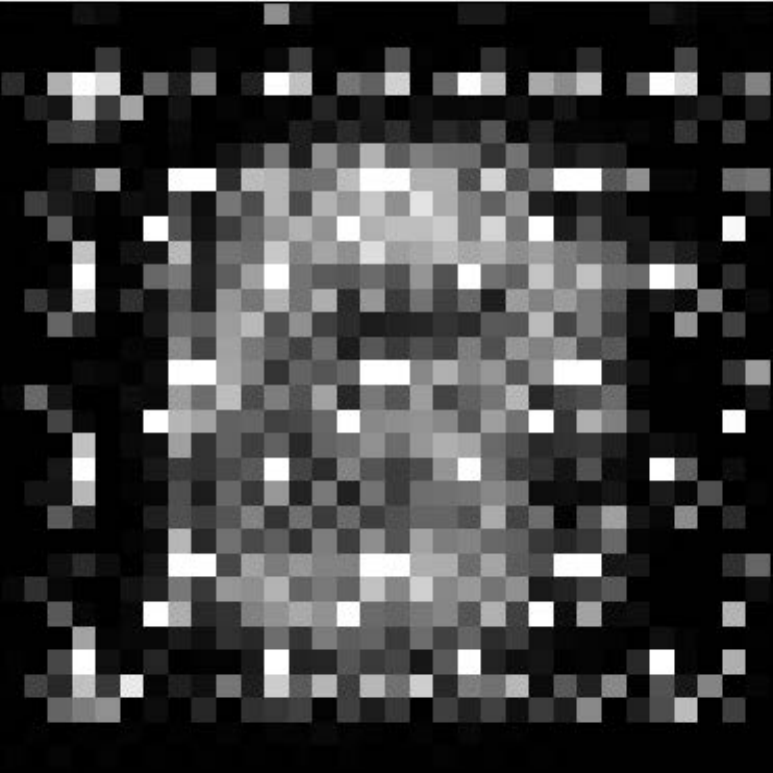}
& \includegraphics[width=0.06\textwidth]{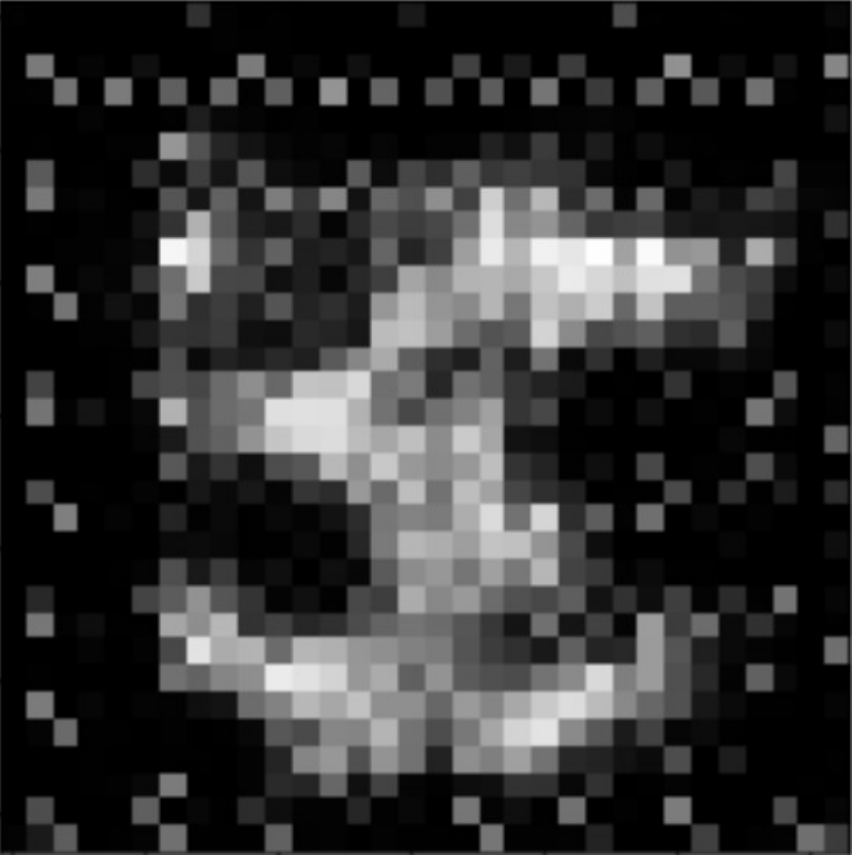} \\
\bottomrule
\end{tabular}
\label{tab-ablation}
\end{table*}

\section{Discussion}\label{sec-discussion}

Unlike traditional inference attacks~\cite{zhu2020deep, luo2020feature} that focus on inferring privacy at the individual record level, the GAN-based attack~\cite{hitaj2017deep} targets the recovery of group-level information (data distribution) in private datasets.
While DP~\cite{dp-sgd, xiao2010differential} offers robust protection against record-level attacks, the development of defense mechanisms with theoretical guarantees against group-level attacks is still limited.
One approach to protect group-level information is to extend traditional DP mechanisms to group-level settings, as described in \cite{zhang2022understanding,jiang2024protecting}. However, this approach may result in significant accuracy degradation for the model. For instance, training a federated classifier on Cifar-10 with a privacy budget of $\epsilon=1.5$ can lead to up to $16.99\%$ accuracy degradation~\cite{zhang2022understanding}.
The research on defense mechanisms that can simultaneously defend against group-level attacks while preserving model utility is still limited. Our work represents the first study to explore defense mechanisms for protecting group-level information while minimizing model performance degradation.
The analysis of the theoretical privacy guarantee of \scheme is challenging due to its objective of obfuscating the visual distribution of the training images $\hat{X}$. Because, unlike pixel values, there are currently no widely recognized metrics that can precisely quantify the visual indistinguishability between images restored by GAN-based attacks and the original images. Additionally, the level of indistinguishability of visual features is subjective and may vary among different human observers.
Nevertheless, our study represents an initial step in the development of defense mechanisms for protecting group-level information. While \scheme is an empirical study, our experimental results demonstrate its effectiveness in defending against GAN-based attacks compared to traditional DP mechanisms, with minimal impact on model accuracy, typically below $5\%$.

{\color{black}

\section{Related Work}\label{sec-related-work}

% \noindent
% \textbf{Inference Attacks on Machine Learning.}
\subsection{{\color{black}Inference Attacks on Machine Learning}}
Machine Learning (ML) algorithms are vulnerable to various forms of inference attacks, including membership inference~\cite{shokri2017membership,salem2018ml,nasr2019comprehensive, melis2019exploiting}, property inference~\cite{propertyFL, propertyPermutation, propertyPoison, melis2019exploiting,jiang2024distribution,luodiffusion}, and feature inference (also known as model inversion)~\cite{fredrikson2015model, luo2020feature, wu2016methodology,zhu2023passive, zhu2020deep,hitaj2017deep,wang2019beyond,luo2022feature,jayaraman2022attribute}. 
{\color{black}\textit{Membership inference}} attacks determine whether a specific sample is part of the training dataset. \cite{shokri2017membership} introduces a shadow model training technique for membership inference, which is later improved by \cite{salem2018ml} under loose assumptions, and by \cite{nasr2019comprehensive} in the setting of federated learning.
{\color{black}\textit{Property inference}} attacks aim to infer aggregate statistics of the training dataset, such as the statistics of software execution traces~\cite{propertyPermutation} and the age distribution of a bank's customer data~\cite{jiang2024distribution}.
{\color{black}\textit{Feature inference}} attacks attempt to reconstruct the original inputs from ML applications. \cite{fredrikson2015model} explores model inversion attacks in decision trees and neural networks using labels and black-box access to the target model. \cite{luo2020feature} proposes to infer the training records in federated learning via model predictions. \cite{zhu2020deep} restores the private training data using only model gradients.
In federated learning, recent studies exploit {\color{black}\textit{GAN-based feature inference attacks}}.
\cite{hitaj2017deep} proposes a GAN-based threat model where an adversary tries to extract representatives of a class not owned by him.
By taking the shared global model as the discriminator of GAN, the adversary can further mislabel the pseudo-samples generated by the generator to trick other participants into leaking more private information.
\cite{wang2019beyond} assumes a malicious server and utilizes GANs to explore user-level privacy leakage, similar to the method in \cite{hitaj2017deep}.
Note that in the literature, GANs have also been used in other types of attacks, such as membership inference attacks~\cite{chen2020gan-membership,lin2021gan-membership-privacy}, backdoor attacks~\cite{mei2023gan-backdoor-privacy,hu2022gan-backdoor-generating}, and adversarial attacks~\cite{zhu2023gan-adversarial-frequency,na2022gan-adversarial-unrestricted}.

In this paper, we focus on {GAN-based feature inference attacks}.
The main difference between GAN-based feature inference attacks~\cite{hitaj2017deep,wang2019beyond} and traditional feature inference attacks~\cite{fredrikson2015model, luo2020feature, wu2016methodology, zhu2020deep,jayaraman2022attribute} is that the former focuses on inferring class representations (i.e., group-level information) via a GAN model from model gradients, whereas the latter aims to restore record-level private information (i.e., individual model inputs) from publicly released information, such as data labels~\cite{fredrikson2015model}, model predictions~\cite{luo2020feature}, and gradients~\cite{zhu2020deep}.
Record-level feature inference attacks can be effectively mitigated by state-of-the-art defense mechanisms in FL, such as  dropout~\cite{melis2019exploiting,salem2018ml}, differential privacy~\cite{shokri2015privacy,bagdasaryan2018backdoor}, and secure multiparty computation~\cite{bonawitz2017practical}. However, GAN-based feature inference attacks pose significant challenges for mitigation. 
The effectiveness of these attacks is closely tied to the practicality of the federated model~\cite{hitaj2017deep}, making them difficult to defend against. To the best of our knowledge, there are no defense mechanisms that can effectively counter GAN-based attacks without causing substantial performance degradation to the federated model.
In this paper, we focus on developing defenses against GAN-based feature inference attacks~\cite{hitaj2017deep, wang2019beyond} that can achieve a favorable privacy-utility trade-off.

% \vspace{0.8mm}
% \noindent
% \textbf{Defenses against Inference Attacks.}
\subsection{\color{black}Defenses against Inference Attacks}
Numerous defense methods have been proposed to mitigate the risks incurred by inference attacks, including dropout~\cite{melis2019exploiting,salem2018ml,scheliga2023dropout}, model stacking~\cite{salem2018ml}, anomaly detection~\cite{bagdasaryan2018backdoor}, differential privacy~\cite{shokri2015privacy,bagdasaryan2018backdoor,geyer2017differentially,jiang2024protecting,guo2022bounding}, secure multiparty computation~\cite{wu2020privacy, Mohassel17,bonawitz2017practical,zhang2020batchcrypt,truex2019hybrid}, and InstaHide~\cite{huang2020instahide}.
{\color{black}\textit{Dropout}}~\cite{melis2019exploiting,salem2018ml,scheliga2023dropout} and {\color{black}\textit{model stacking}}~\cite{salem2018ml} are basically regularization methods, which are mainly used to prevent model overfitting, thereby reducing the amount of information memorized by ML models. {\color{black}\textit{Anomaly detection}}~\cite{bagdasaryan2018backdoor} is utilized to mitigate poisoning-based attacks performed by active adversaries. 
Traditional {\color{black}\textit{differential privacy}} methods~\cite{shokri2015privacy,bagdasaryan2018backdoor,guo2022bounding} provide rigorous theoretical guarantees for privacy protection, but they are applicable to individual-level attacks instead of group-level attacks. Although recent studies~\cite{geyer2017differentially,jiang2024protecting,jiang2024calibrating} have extended DP guarantees to group-level privacy, these methods can significantly degrade model performance.
{\color{black}\textit{Secure multiparty computation}}~\cite{wu2020privacy, Mohassel17,bonawitz2017practical} can impose substantial computational costs during model training and has not been widely applied in real-world applications.

Among these defense methods, {\color{black}\textit{InstaHide}}~\cite{huang2020instahide} is most similar to our approach, i.e., using {Mixup}~\cite{zhang2017mixup} to corrupt the visual features of private training images. However, there are three main differences between InstaHide and \schemens.
First, InstaHide mixes multiple real images into one training image, while our scheme uses a GAN model to generate multiple fake images and then mixes these fake images with real images.
Second, InstaHide obfuscates image visual features by randomly flipping pixel signs, leading to uncontrollable model performance degradation. In contrast, \scheme obfuscates visual features using a specifically designed loss function, allowing the privacy-utility tradeoff to be tuned for different application scenarios.
Third, \cite{carlini2020attack} and \cite{luo2021fusion} have experimentally demonstrated that private images can be reconstructed from the mixup images generated by InstaHide on the condition that the same private image is mixed into multiple training images. Our scheme mitigates this risk by ensuring that each private image is mixed only once.
It is important to note that none of the aforementioned defenses, including InstaHide, are specifically designed to counter GAN-based attacks. To the best of our knowledge, this is the \textit{first} paper to exploit defenses against GAN-based feature inference attacks in the context of federated learning.

}
\section{Conclusion}\label{sec-conclusion}
{\color{black}
In this paper, we introduce \scheme, a novel approach designed to protect the privacy of participants' training data against GAN-based feature inference attacks in the context of federated learning. By strategically distorting the original distribution of the private training dataset, \scheme effectively prevents attackers from generating distinguishable images while maintaining a high-performance federated classifier. 
Our experimental results demonstrate that, compared to the widely adopted DP-SGD mechanism, which can degrade model accuracy by up to $12\%$ and allow for a similarity score of up to 0.53 for image reconstruction, our method limits model accuracy degradation to a maximum of $5\%$ and reduces the similarity scores of reconstructed images to at most 0.30. In addition, our experiments show that \scheme can not only defend against GAN-based feature inference attacks but also provide protection against gradient-based record-level reconstruction attacks.
This study represents an initial step toward developing defense mechanisms against GAN-based attacks that can achieve a favorable privacy-utility tradeoff, and we believe it can contribute valuable insights into the advancement of privacy-preserving mechanisms in the context of federated learning.

}

\bibliographystyle{ACM-Reference-Format}
\bibliography{acml21.bib}

\newpage
\appendix
\section{Appendix}

\subsection{Detailed Model Structures}\label{appendix-net-arch}
In this section, We introduce the details of the GAN models and the federated classifier.
Let $N_{out}$ denote the number of feature maps output by each layer.
For both the \textit{attacker's} and \textit{defender's generator}, we adopt the same architecture, as outlined in Table~\ref{tab-g}. It is worth noting that the input layer's $N_{out}$ corresponds to the concatenation of a Gaussian noise vector with 100 elements and a label embedding with 10 elements.
The architecture of the \textit{defender's discriminator} is presented in Table~\ref{tab-d-d}. Here, $C(x)$ represents a set of 64 feature maps. We augment the label embedding to match the size of $C(x)$ and subsequently concatenate them to serve as the input for the defender's discriminator.
As for the \textit{federated model}, its architecture is provided in Table~\ref{tab-fl-m}. In the GAN-based attack~\cite{hitaj2017deep}, the adversary can construct the discriminator of the attacking GAN by expanding the output dimension of the federated model from 10 to 11. The additional class is utilized to classify the fake images.

\begin{table}[ht!]
\caption{The generator architecture used in both the defender's and attacker's GAN models. $N_{out}$ denotes the number of feature maps output by each layer.}\label{tab-g}
\centering
% \vspace{-2mm}
\begin{tabular}{ | c |l | c | l|}
\toprule
Layer & Name & $N_{out}$ & Function \\
\toprule
0 & Input & 110 & Noises and labels  \\
1 & ConvTranspose2d & 256 & Kernel $4\times 4$ \\
2 & BatchNorm2d & 256 & Batch normalization \\
3 & ReLU & 256 & Activation \\
4 & ConvTranspose2d & 128 & Kernel $4\times 4$ \\
5 & BatchNorm2d & 128 & Batch normalization \\
6 & ReLU & 128 & Activation \\
7 & ConvTranspose2d & 64 & Kernel $4\times 4$ \\
8 & BatchNorm2d & 64 & Batch normalization \\
9 & ReLU & 64 & Activation \\
10 & ConvTranspose2d & 3 & Kernel $4\times 4$ \\
11 & Tanh & 3 & Activation \\
\bottomrule
\end{tabular}
% \vspace{-3mm}
\end{table}

\begin{table}[ht!]
\caption{The discriminator architecture used in the defender's GAN model. The negative slope used in leaky ReLU is 0.2.}\label{tab-d-d}
\centering
% \vspace{-2mm}
\begin{tabular}{ | c |l | c | l|}
\toprule
Layer & Name & $N_{out}$ & Function \\
\toprule
0 & Input & 128 & $C(x)$ and labels  \\
1 & Conv2d & 128 & Kernel $4\times 4$ \\
2 & InstanceNorm2d & 128 & Instance normalization \\
3 & LeakyReLU & 128 & Activation \\
4 & Conv2d & 256 & Kernel $4\times 4$ \\
5 & InstanceNorm2d & 256 & Instance normalization \\
6 & LeakyReLU & 256 & Activation \\
7 & Conv2d & 512 & Kernel $4\times 4$ \\
8 & InstanceNorm2d & 512 & Instance normalization \\
9 & LeakyReLU & 512 & Activation \\
10 & Conv2d & 1 & Kernel $4\times 4$ \\
\bottomrule
\end{tabular}
% \vspace{-3mm}
\end{table}

\begin{table*}[ht!]
\caption{The architecture of the federated model. By changing $N_{out}$ of  the final layer from 10 to 11 (the additional class is for fake images), we can use the federated model as the discriminator of the attacker's GAN model~\cite{hitaj2017deep}.}\label{tab-fl-m}
\centering
% \vspace{-2mm}
\begin{tabular}{ | c |l | c | l|}
\toprule
Layer & Name & $N_{out}$ & Function \\
\toprule
0 & Input & 3 & Input image  \\

1 & Conv2d & 16 & Kernel $3\times 3$ \\
2 & ReLU & 16 &  Activation\\
3 & MaxPool2d & 16 & Kernel $2\times 2$ \\

4 & Conv2d & 64 & Kernel $3\times 3$ \\
5 & ReLU & 64 &  Activation\\
6 & MaxPool2d & 64 & Kernel $2\times 2$ \\

7 & Dropout & 64 & Rate 0.5 \\

8 & Linear & 100 & Dense Layer \\
9 & ReLU & 100 &  Activation\\
10 & Linear & 10 & Dense Layer \\
\bottomrule
\end{tabular}
% \vspace{-3mm}
\end{table*}

\end{document}